\documentclass[letterpaper,onecolumn]{IEEEtran}
\ifCLASSOPTIONcompsoc
\else
\fi
\usepackage{graphicx}
\usepackage{url}
\usepackage{cite}
\usepackage{caption}
\usepackage{epstopdf}
\usepackage{subcaption}
\usepackage{bytefield}
\usepackage{flushend}

\usepackage[rgb]{xcolor}
\usepackage[author={Arsany and Raymond}]{pdfcomment}
\usepackage[capitalise]{cleveref}
\ifCLASSINFOpdf
\else
\fi

\def \sys {\textit{PUNCH}}
\begin{document}
\title{Primary User-aware Network Coding for Multi-hop Cognitive Radio Networks}

\author{
  \IEEEauthorblockN{Arsany Guirguis$^\dag$, Raymond Guirguis$^\dag$, Moustafa Youssef$^\ddag$}\\
  \IEEEauthorblockA{
    $^\dag$Dept. of Computer and Systems Engineering\\
    Faculty of Engineering\\
    Alexandria University, Egypt\\
    \{arsany.hany,raymond.milad\}@alex.edu.eg
  }
  \\
  \and
  \IEEEauthorblockA{
    $^\ddag$Wireless Research Center\\
    Alexandria University and E-JUST\\
    Alexandria, Egypt\\
    moustafa.youssef@ejust.edu.eg\\
  }}

\IEEEcompsoctitleabstractindextext{%
\begin{abstract}
Network coding has proved its efficiency in increasing the network performance for traditional ad-hoc networks. In this paper, we investigate using network coding for enhancing the throughput of multi-hop cognitive radio networks. We formulate the network coding throughput maximization problem as a graph theory problem, where different constraints and primary users' characteristics are mapped to the graph structure. We then show that the optimal solution to this problem in NP-hard and propose a heuristic algorithm to efficiently solve it.

Evaluation of the proposed algorithm through NS2 simulations shows that we can increase the throughput of the constrained secondary users' network by 150\% to 200\% for a wide range of scenarios covering different primary users' densities, traffic loads, and spectrum availability.

\end{abstract}

}

\maketitle

\IEEEdisplaynotcompsoctitleabstractindextext

\IEEEpeerreviewmaketitle

\section{Introduction}

With the wide spread use of mobile devices, wireless networks have become indispensable to support always-on anywhere connectivity. Cognitive Radio Networks (CRNs) emerged as a paradigm to solve the problem of limited spectrum availability and the inefficiency in the spectrum usage. In CRNs, secondary users (SUs) are allowed to access the spectrum as long as they do not interfere with primary users (PUs) who have the license of the band and have the higher priority of using it. However, the wireless spectrum is scarce and the demand keeps increasing with the growing demand of bandwidth-intensive applications, such as video streaming, that require new solutions for better spectrum utilization.

\begin{figure}[!t]
\centering
	\begin{subfigure}[t]{0.5\textwidth}
	\centering
    \includegraphics[width=3.1in]{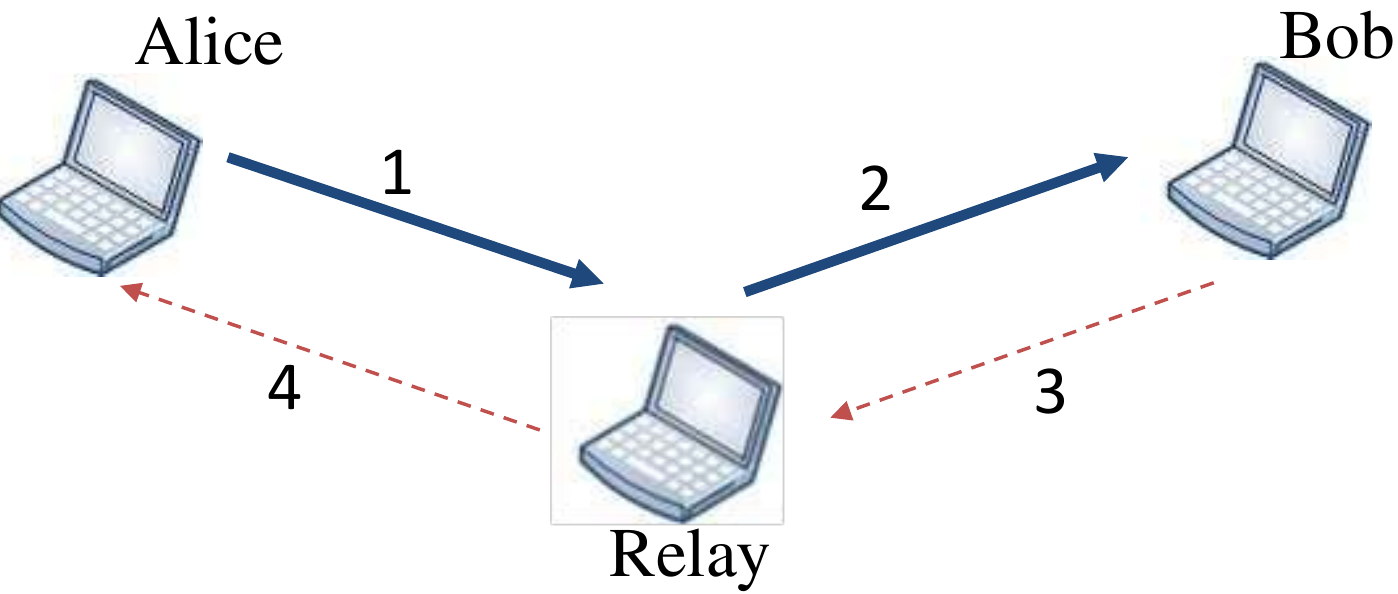}
	\caption{Without network coding.}
	\label{fig:intro_raw}
	\end{subfigure}
	\begin{subfigure}[t]{0.5\textwidth}
	\centering
	\includegraphics[width=3.1in]{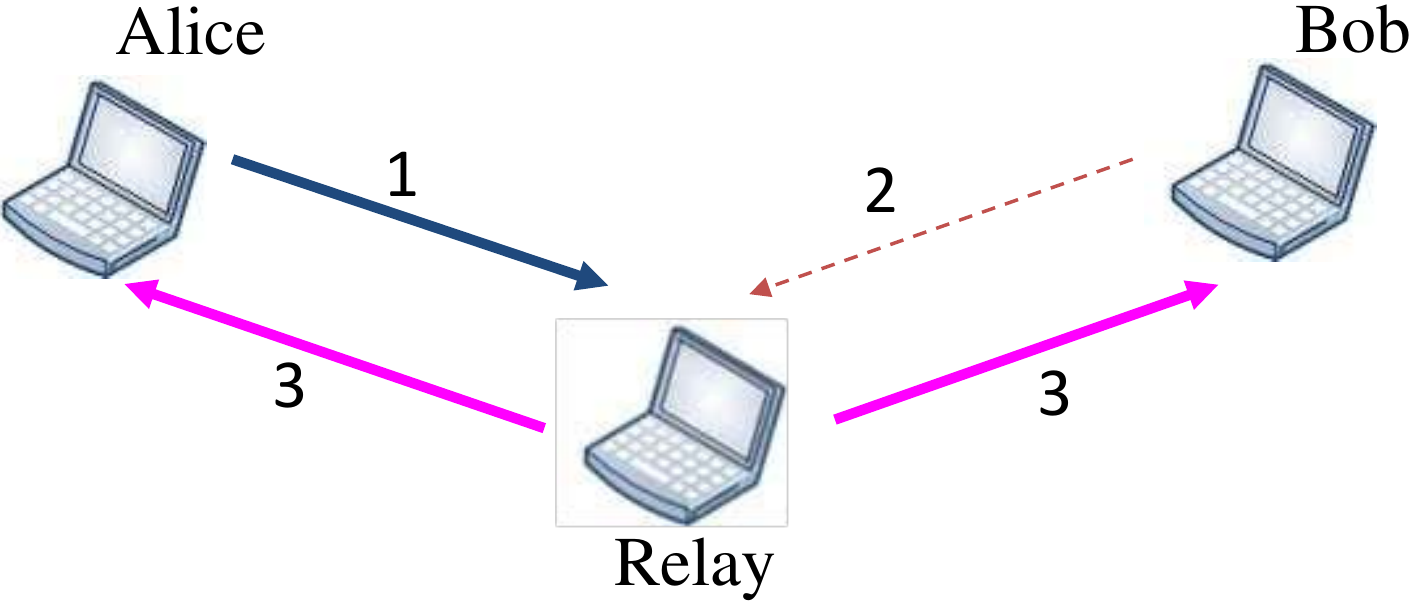}
	\caption{With network coding.}
	\label{fig:intro_denoised}
	\end{subfigure}
\caption{Example that shows how network coding can increase throughput. Numbers on arrows represent the order of transmissions.}
\label{AliceAndBob}
\end{figure}

Since the pioneering work in \cite{net-coding_first}, network coding has proved its ability to increase the utilization of traditional wireless networks \cite{xors,netcoding,energy,Fragouli:2007:NCA:1345035.1345036,li2003linear,Deb05networkcoding,Koetter:2003:AAN:948928.948936}. It takes advantage of the broadcast nature of wireless networks to allow intermediate nodes to efficiently combine packets before forwarding. This way, the information content in each transmission increases by forwarding multiple packets in a single transmission. For example, in \cref{AliceAndBob} Alice and Bob are two users who want to exchange a pair of packets through a relay node. Without network coding (Figure 1a), Alice sends her packet to the relay and so does Bob. Then the relay forwards Alice's packet to Bob and Bob's one to Alice. This process requires four transmissions in total. Using network coding (Figure 1b), Alice and Bob send their packets to the relay which XORs the two packets and broadcasts the XOR-ed version. Alice and Bob receive the XOR-ed packet and can extract each other’s packet by XOR-ing again with their own packets. This process takes only three transmissions, leading to increased throughput for the network coding case.

Network Coding was studied in the context of traditional wireless  ad-hoc networks to increase throughput (e.g. \cite{xors}), reduce retransmissions (e.g. \cite{netcoding}), or for energy-efficient broadcast (e.g. \cite{energy}). Extending network coding, however, to the case of CRNs is not straightforward. In particular, the PUs' activity and location should be taken into account to determine the best packets combination that maximizes the coding gain.

There has been some work in leveraging network coding in CRNs \cite{pucoding,multicast, geographical}. \cite{pucoding} uses network coding in the \textbf{primary users' network}, rather than the secondary users network. The intuition is that increasing the efficiency of the PUs' network can release more spectrum for use by the SUs' network. However, this requires changes to the PUs' network, where PUs have no incentive for doing this for the sake of SUs. On the other hand, the work in \cite{multicast} focuses on multi-cast traffic, which is not the common traffic pattern in wireless networks. Finally, the work in \cite{geographical} proposes a geographical opportunistic routing scheme combined with network coding for CRNs. However, they only handle PUs' activities in determining the routes, not in the network coding process, and do not take transmission impairments into account. In addition, they focus on minimizing the average number of re-transmissions per packet, and not maximizing throughput.

In this paper, we present the PUNCH algorithm for \textbf{PU}-aware \textbf{N}etwork \textbf{C}oding in multi-\textbf{h}op CRNs. Compared to related work, our algorithm targets the constrained secondary users' network unicast traffic, aims at maximizing throughput, and takes both the PUs and the transmissions impairments into account. We start by formalizing the throughput maximization problem as a graph theoretic problem, where the different constraints are mapped to the graph structure. We then show that the optimal coding for throughput maximization is an NP-hard problem. Therefore, we present a heuristic for solving it efficiently. Practical considerations such as reliable transmission are also presented.

Evaluation of the proposed algorithm using NS2 \cite{ns2} simulations shows significant increase in throughput up to 200\% compared to protocols that do not leverage network coding. This increase is maintained under various network scenarios.

The rest of the paper is structured as follows. In \cref{a}, we present an overview of \sys{} operation and the system model. In \cref{b}, we present the system details. In \cref{c}, we present the details of our implementations. In \cref{d}, we evaluate the proposed algorithm using NS2 simulations. Finally, we conclude the paper in \cref{e} and give directions for future work.

\section{Overview and System Model}\label{a}
Before going into details of our proposed algorithm in the next section, we first introduce the main terminology used in the paper, followed by the system model, and finally give the main concepts of operation.

\subsection{Terminology}
We use the following terminology:
\begin{itemize}
\item Native packet: A single original packet as sent from its source without being XOR-ed with other packets.
\item Encoded or XOR-ed packet: a packet that results from encoding two or more native packets.
\item Output queue: Each node has its First In First Out FIFO queue where it keeps the packets it received from its application layer or from its neighbours to be forwarded.
\item Packet pool: Each node keeps packets it heard for a certain amount of time to use them in decoding XOR-ed packets.
\item Reception reports: Information that each node broadcasts periodically to inform its neighbours of packets it has in its packet pool.
\end{itemize}

\subsection{System Model}
We consider a CRN where PUs and SUs co-exist. PUs are the primary holder of the spectrum access rights and have priority over SUs, i.e. SUs' traffic should not affect the operation of the primary network. For modeling the PUs' activities, we adopt the commonly used two-state ON-OFF birth-death process model, where the PU can either be active (ON) or inactive (OFF). The dwell time in each state follows an exponential distribution with parameters $\lambda$ and $\mu$ for the active and inactive states respectively.

We also assume that the number of PUs in range can be estimated by a SU based on
passive sensing. Also their distribution parameters are assumed
to  be  known  through  estimation.  We  assume  that  PUs  are
stationary. This is common in many CRN scenarios such as
white space-based CRNs.
We also do not make a specific assumption on the MAC or higher layer protocols for the PUs' system.

Finally, we assume that there is a routing protocol that is running in the background to discover routes to the different destinations \cite{crn_survey,prob_wcnc14,DZP13,DZP_ARXIV,Islam11}. Our goal is to code the packets along the chosen routes, for  unicast traffic between different source-destination pairs, to maximize the secondary network throughout given the different link qualities to neighbours  and the PUs' activities. Linear network coding \cite{li2003linear} is used as well as decoding is performed every hop, as compared to end-to-end, to increase the practicality of the algorithm.

\subsection{Concepts of Operation}
As in standard network coding, our algorithm is based on three main concepts of operation:

{\bf (a) Opportunistic Listening:} The shared medium in wireless networks introduces the advantage of overhearing, where a node can hear any packet within its transmission region even if it is not destined to it. This can be used to decode other encoded packets and increases the opportunities of successful encoding and decoding of packets at the sender and receiver.

{\bf (b) Opportunistic Coding:} Whenever a node wants to transmit a packet, it examines that packets currently in its output queue and decides whether there is an opportunity to encode multiple packets together. To maximize the throughput, the set of packets that maximize the number of neighbours which can decode the encoded packet is the best option. In other words we want to maximize the number of native packets delivered in a single transmission. This decision  should be a function of the ability of the neighbours to decode the packets, PUs' activity, as well as the quality of the link between the sender and the receiver.

{\bf (c) Learning Neighbour State:} We assume that each node has information about packets at its neighbours. This can be done by overhearing the packets transmitted between neighbours. To further enhance the accuracy of estimating the packets at the neighbours, each node also periodically broadcasts reception reports to its neighbours informing them of the packets it has in its packet pool.

\section{PUNCH Details}\label{b}
We start by formulating the forwarding throughput maximization in multi-hop CRNs using network coding problem as a graph theory problem. We then discuss how to change the graph parameters to fit different PUs and link quality constraints.
\begin{figure*}[!t]
\centering
	\begin{subfigure}[t]{0.3\textwidth}
	\centering
    \includegraphics[width=2.1in]{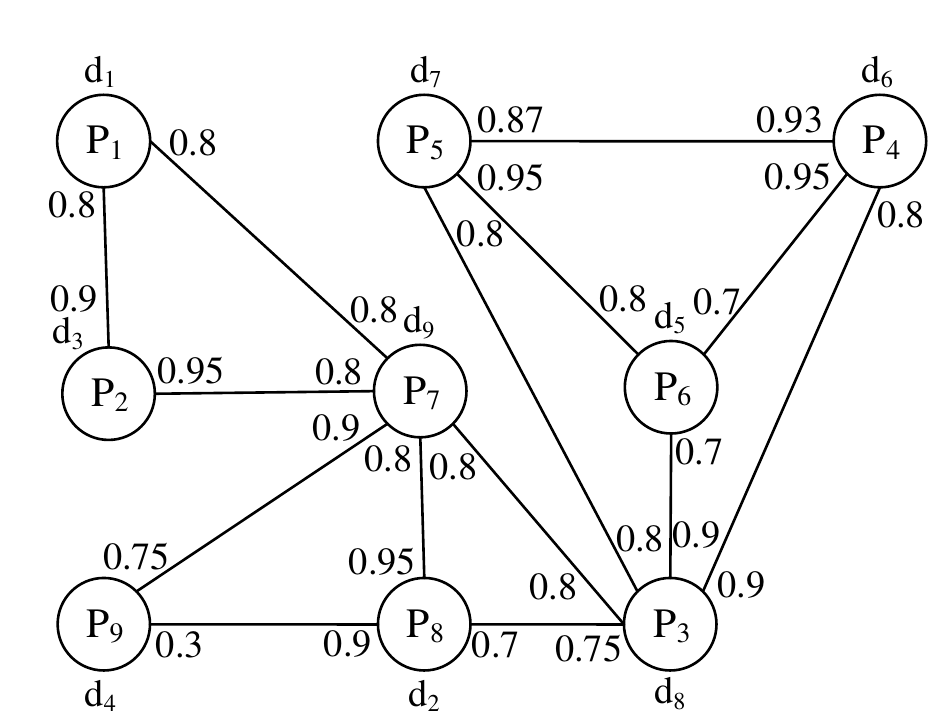}
	\caption{Coding graph: Vertices represent the packets in the output queue (tagged with the destination ID). Weights represent the probability for one destination to decode packets if these both packets (connected by this edge) are encoded together.}
	\label{directed}
	\end{subfigure}
	\begin{subfigure}[t]{0.3\textwidth}
	\centering
	\includegraphics[width=2.1in]{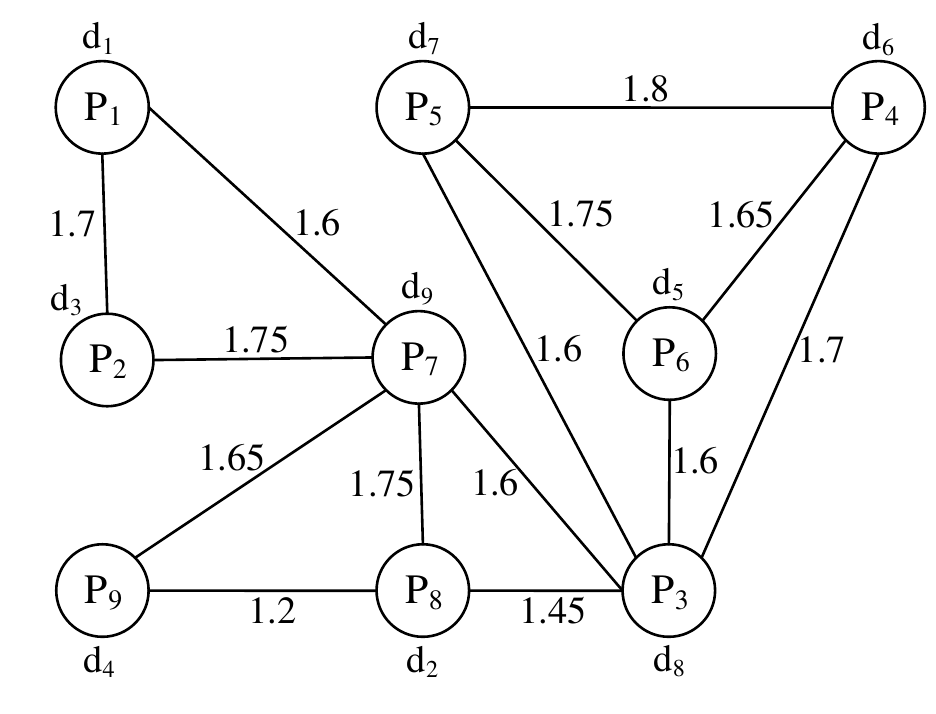}
	\caption{Coding gain graph: Undirected graph obtained from the coding graph. Each edge weight represents the expected coding gain by encoding the two packets represented by the edge.}
	\label{undirected}
	\end{subfigure}
	\begin{subfigure}[t]{0.3\textwidth}
	\centering
	\includegraphics[width=2.1in]{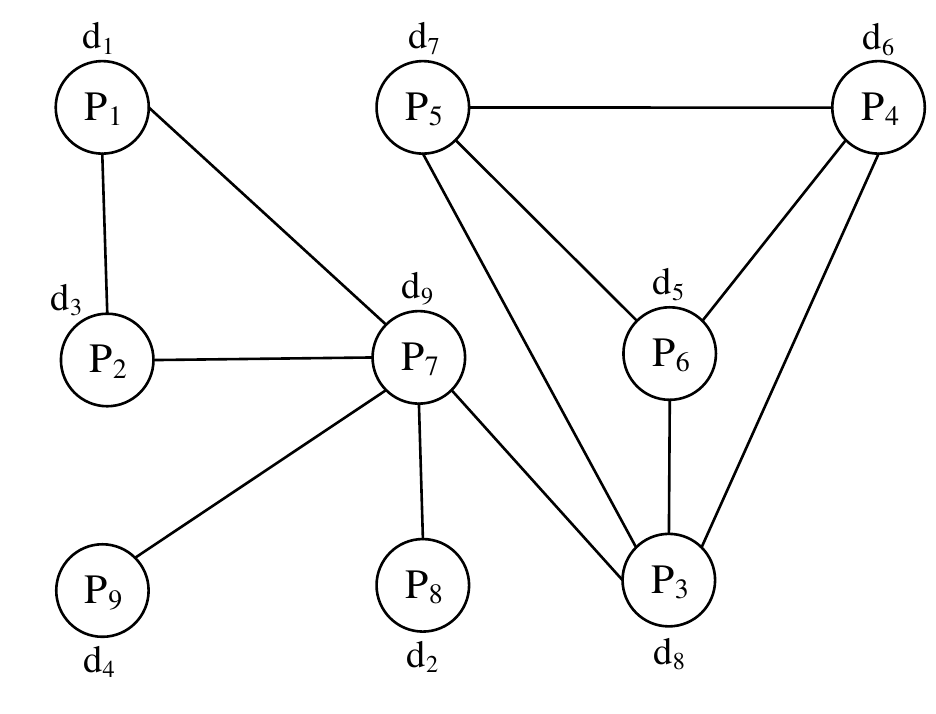}
	\caption{Final un-weighted graph: Obtained from the coding gain graph by a threshold on the edge weight (1.5 in this case). This represents the minimum quality required for coding.}
	\label{unweighted}
	\end{subfigure}
\caption{Example on the different graphs used to obtain the best encoding at a given node.}
\label{fig:graphs}
\end{figure*}

\subsection{Problem Formulation}
Define the {\bf coding graph} $G=(V, E)$ as the weighted directed graph constructed at each node to represent the different coding opportunities for the packets in the output queue of the node on a specific channel. Each vertex in the graph represents a packet in the output queue (waiting to be sent). The vertex is tagged with the ID of the neighbour this packet is destined to. This destination is known from the routing protocol. An edge in the coding graph represents a coding opportunity between the two packets at the end of the edge. The weights ($w_{ij}$) on the edges connecting nodes $i$ and $j$ represent the probability that the two encoded packets the edge represents can be decoded at each neighbour. This weight is a function of the availability of the coded packets at the neighbour, PUs' activity, and channel quality between the sender and receiver. Therefore it is not symmetric with respect to the two neighbours that will receive the encoded packet represented by the edge.

We then convert the coding graph to the \textbf{coding gain graph} that represents the coding gain achieved by coding each packet. In particular, the coding gain graph is an undirected version of the coding graph whose edge weights equal $w_{ij}+w_{ji}$. This new edge weight is directly proportional to the expected number of packets to be delivered in a single transmission as compared to only one packet in the no-coding case.

Note that any clique within this graph represents the coding of all packets within this clique. Therefore, in order to maximize the network throughput, one needs to select the maximum weighted clique within the graph. In other words, the clique with the maximum sum of the weights  of its edges should lead to the maximum coding gain, taking into account the PUs' activities and link qualities.

\subsection{Solving for the Optimal Coding}
The maximum edge-weighted clique problem can be shown to be NP-hard by polynomial reduction from the  max-clique problem \cite{clique}. To efficiently solve this problem, we use a heuristic to convert it to an unweighed graph by putting a threshold on the edge weights (taken as 1.5 in our case), representing the minimum acceptable coding gain. Any edge whose weight falls below this threshold is discarded. We then apply an efficient greedy heuristic to solve the max-clique problem on the resulting un-directed graph \cite{Bomze99themaximum}. For more details and example, please refer to our technical report [ref].

\subsection{Example}
Figure \ref{directed} shows an example for the coding graph for one of the nodes. The node has nine packets in its output queue. Therefore there are nine vertices in the graph. Each vertex is tagged with the destination node this packet is destined to. The weights on the edges represent the probability to be decoded by the other destination if combined with the other packet.

Figure \ref{undirected} shows the corresponding coding gain graph, where the weight of each undirected link represents the expected gain of using this encoding. Finally, Figure \ref{unweighted} shows the thresholded graph, where the edges whose weight falls below 1.5 are discarded from the coding gain graph. The max-clique heuristic is applied to this final graph to determine the best encoding, $\{P3, P4, P5, P6\}$ is this case.

\subsection{Graph Construction}

There are a number of considerations that need to be taken into account when constructing the coding graph. We discuss them in this section including how the edge weights are calculated.

First, one should never code packets that are forwarded to the same neighbour as both packets are new to the neighbours and, therefore, cannot be decoded. To avoid this, we do not add any edge to the coding graph between packets destined to the same node. This is achieved by checking the vertex destination tag.

\newcommand{\colorbitbox}[3]{%
\rlap{\bitbox{#2}{\color{#1}\rule{\width}{\height}}}%
\bitbox{#2}{#3}}

\begin{figure}[!t]
  \centering
\definecolor{lightgray}{gray}{0.8}
  \begin{bytefield}[bitheight=1em]{32}

\wordbox{1}{MAC Header}\\
    \begin{rightwordgroup}{\rotatebox{90}{\small XORed} \rotatebox{90}{\small Packets}}
	
\colorbitbox{lightgray}{32}{Number of encoded packets}\\

      \bitbox{16}{Packet ID} &
      \bitbox{16}{Next Hop}\\
	\wordbox[blr]{1}{$\vdots$\\ [1ex]}

    \end{rightwordgroup} \\

  \begin{rightwordgroup}{\rotatebox{90}{\small Reception} \rotatebox{90}{\small Report}}

\colorbitbox{lightgray}{32}{Number of packets in reception report}\\

      \bitbox{16}{Packet ID} &
      \bitbox{16}{Packet ID} \\
      \bitbox[blr]{16}{$\vdots$ \\[1ex]} &
      \bitbox[blr]{16}{$\vdots$ \\[1ex]}

  \end{rightwordgroup}\\

  \begin{rightwordgroup}{\rotatebox{90}{\small Ack} \rotatebox{90}{\small Block}}

\colorbitbox{lightgray}{32}{Number of acked packets}\\

      \bitbox{16}{Packet ID} &
      \bitbox{16}{Packet ID} \\
      \bitbox[blr]{16}{$\vdots$ \\[1ex]} &
      \bitbox[blr]{16}{$\vdots$ \\[1ex]}

  \end{rightwordgroup}\\
  \begin{rightwordgroup}{\rotatebox{90}{\small PU} \rotatebox{90}{\small Block}}

\colorbitbox{lightgray}{32}{Number of PUs}\\

      \bitbox{16}{PU index} &
      \bitbox{16}{PU activity ($\lambda_i$)} \\
      \bitbox[blr]{16}{$\vdots$ \\[1ex]} &
      \bitbox[blr]{16}{$\vdots$ \\[1ex]}

  \end{rightwordgroup}\\
  \begin{rightwordgroup}{\rotatebox{90}{\small Link Qual.} \rotatebox{90}{\small Block}}

\colorbitbox{lightgray}{32}{Number of links}\\

      \bitbox{16}{Neighbour ID} &
      \bitbox{16}{$P_\textrm{link}$} \\
      \bitbox[blr]{16}{$\vdots$ \\[1ex]} &
      \bitbox[blr]{16}{$\vdots$ \\[1ex]}

  \end{rightwordgroup}\\

\wordbox{1}{IP Header}
  \end{bytefield}
  \caption{\sys{} header format.}
  \label{fig:header}
\end{figure}

Second, we want to ensure that all next hops that will receive the encoded packet can decode it.
 This is captured by the edge weight ($w_{ij}$) which represents the probability that the neighbour will correctly decode the packet. Three factors affect the calculation of $w_{ij}$:
\begin{enumerate}
  \item \textbf{Link quality:} A dropped packet due to transmission error will not be received and hence not decoded at the receiver, wasting the bandwidth and reducing throughput. To calculate the probability of packet drop ($P_{\textrm{link}}$), we leverage the lower layer information from the data link layer which calculates the percentage of times transmitted packets to a certain destination are received.
   \item \textbf{Availability of the other coded native packets in the receiver's packet pool:} We leverage reception reports that are broadcast periodically from each node to its neighbours to determine whether the packets exist in the receiver's packet pool or not. A reception report contains the ID of the packets in the packet pool at the destination, the PUs heard at the node, as well as the link quality between the node and its neighbours. Due to the transmission impairments, reception reports may be lost. To reduce this effect, we also leverage packets overhearing to detect if the packet reaches a destination or not. For example, if a node overhears a packet destined to a specific node, it can know that this packet will reach the destination node with a probability ($P_{\textrm{dest}}$) that is a function of the link quality between the packet source and destination and the associated PUs' activity received in previous reception reports.

       To add an edge to the coding graph, the other encoded packet should be in the receiver's packet pool. Note that basing the best encoding on finding the max-clique ensures that all neighbours can decode the packet with high probability.
  \item \textbf{PUs' activity:} Given our model for the PUs' activity, the probability ($P_{\textrm{active}}$) that at least one of the $m$ primary users affecting a link will become active during some time period $\tau$ is:
      \begin{equation}
    P_{\textrm{active}} = 1 - e^{- \tau \Sigma_{i=1}^m  \lambda_i }
      \end{equation}
      Where  $ \lambda_i $ represents the parameter of the exponential distribution in the ON period of PU $i$.
Higher values for $\tau$ represents a more stable path.
  \item \textbf{Combining the three terms:} Based on the above factors, the edge weight $w_{ij}$, which is the probability that a packet will be decoded by the destination node, is given by:
       \begin{equation}
        w_{ij}= (1-P_{\textrm{active}}).P_{\textrm{dest}}. (1-P_{\textrm{link}})
      \end{equation}
\end{enumerate}

Finally, other metrics can also be incorporated into the coding graph. For example, to enhance the system delay, one may choose to always encode the packet at the head of the queue. This can be incorporated in the edge weights or in the heuristic algorithm used to solve the max-clique problem.

\section{Implementation details}\label{c}
In this section we present the details of implementing the algorithm.

\subsection{Reliable Transmission}
Since network coding sends one packet to multiple destinations, there is a problem in acknowledging the reception of the packets from all the receivers (the broadcast storm problem \cite{tseng2002broadcast}). To address this problem, we implement a pseudo-broadcast scheme, where we set the destination address of the encoded unicast packet to the address of one of receivers and append a list of all other receivers (whose native packets are encoded in the same packet) in another header that accompany the link-layer header (Figure~\ref{fig:header}). Therefore, the node whose address is used as the packet destination acknowledges the reception using standard link layer mechanisms. Other nodes, whose addresses are in the receivers list, acknowledge the reception in the reception report. The sender node keeps the packets in its output queue (but tags them as already transmitted) till it is acknowledged in the reception report. If the acknowledgement is not received within a certain time, the tag is cleared to allow for retransmission.

\begin{figure}[!t]
\centering
	\begin{subfigure}[t]{0.23\textwidth}
	\centering
    \includegraphics[width=0.9in]{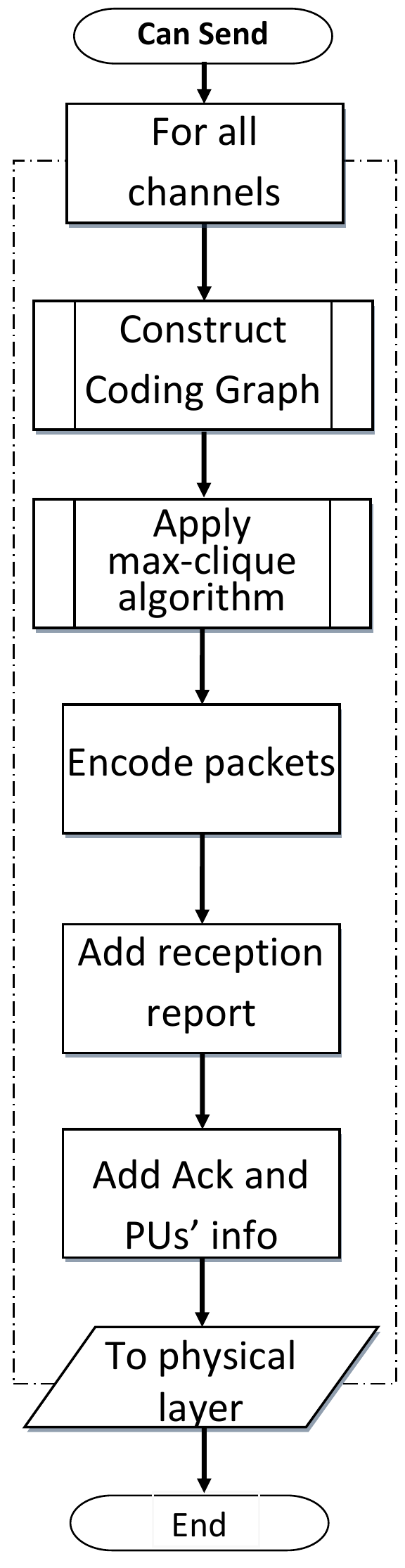}
	\caption{Sender Side}
	\label{fig:intro_raw}
	\end{subfigure}
	\begin{subfigure}[t]{0.23\textwidth}
	\centering
	\includegraphics[width=1.51in]{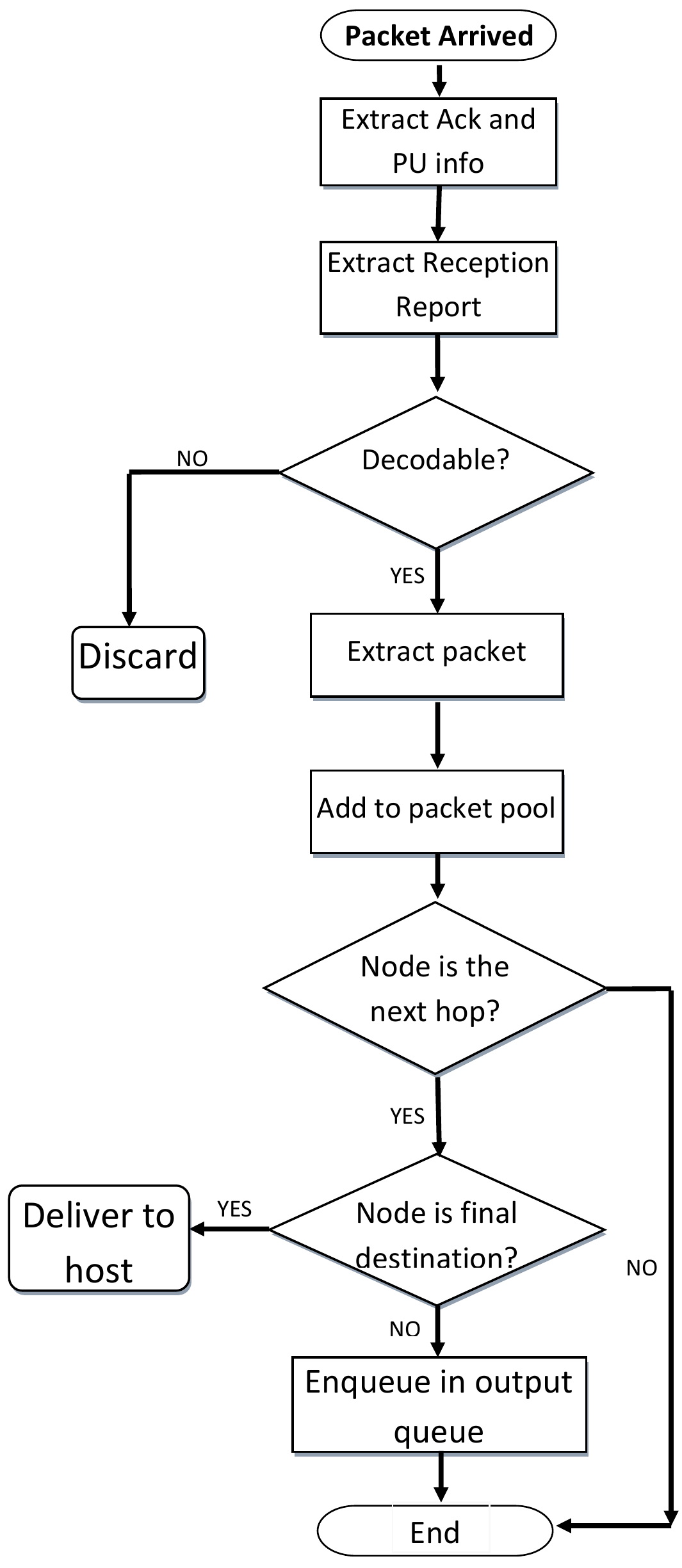}
	\caption{Receiver Side}
	\label{fig:intro_denoised}
	\end{subfigure}
\caption{Flowchart for the coding and decoding operation.}\label{fig:flowchart}
\end{figure}

\subsection{Coding and Decoding}
We implemented \sys{} as a layer between the MAC and network layers. Figure~\ref{fig:flowchart} shows the flowchart of operation. When sending a packet, the sending node constructs/updates its coding graph based on the pending packets in its packet queue. The best encoding is applied to the packets and the XOR-ed packet is forwarded to the physical layer for transmission.

When a node receives an encoded packet, it checks whether it can decode it or not. This is possible only if the node has $n-1$ native packets from the $n$ encoded  packets in its packet pool. The node then checks if this packet is destined to it (from the next hop address in link-layer header). If not, it checks if its address is in the list-of-destinations header.  If it finds itself, it either forwards the packet to the upper layers (if it is the final destination) or adds the packet to its output queue, if it is just a hop to the final destination. Otherwise, it stores this packet in the packet pool as it may help in decoding some other future packets. For detailed flow chart for sending and receiving operations, please refer to our technical report [ref].

\begin{figure}[!t]
\centering
\captionsetup{justification=centering}
\includegraphics[width=2.5in]{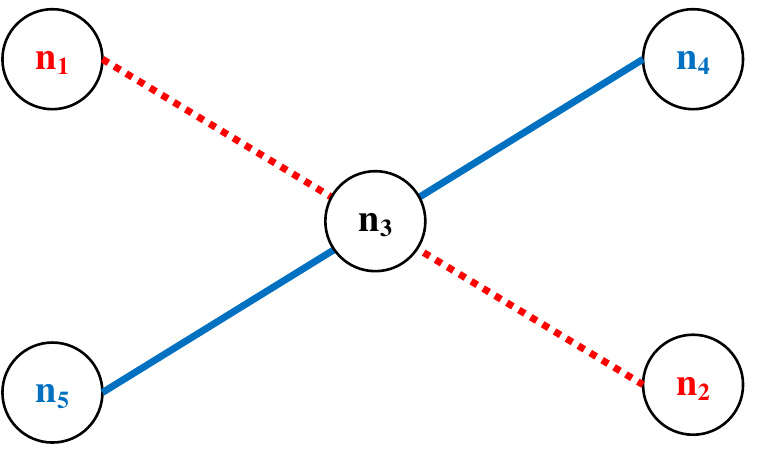}
\caption{The star topology used in evaluating the effect of the different parameters on \sys{}.}
\label{X}
\end{figure}

\begin{figure*}[!t]
\centering
	\begin{subfigure}[t]{0.24\textwidth}
	\centering
    \includegraphics[width=1.8in]{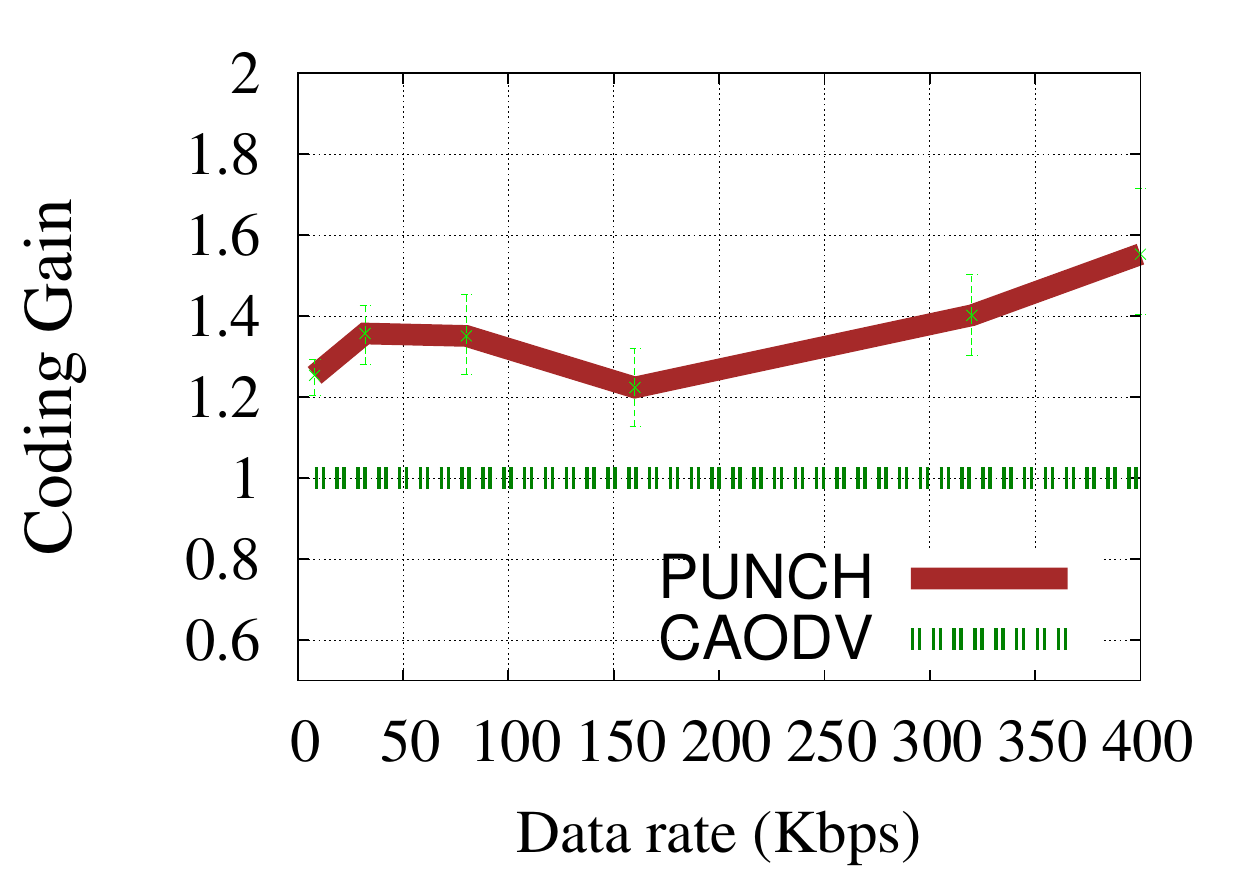}
	\caption{Coding gain.}
	\label{rate_a}
	\end{subfigure}
	\begin{subfigure}[t]{0.24\textwidth}
	\centering
	\includegraphics[width=1.8in]{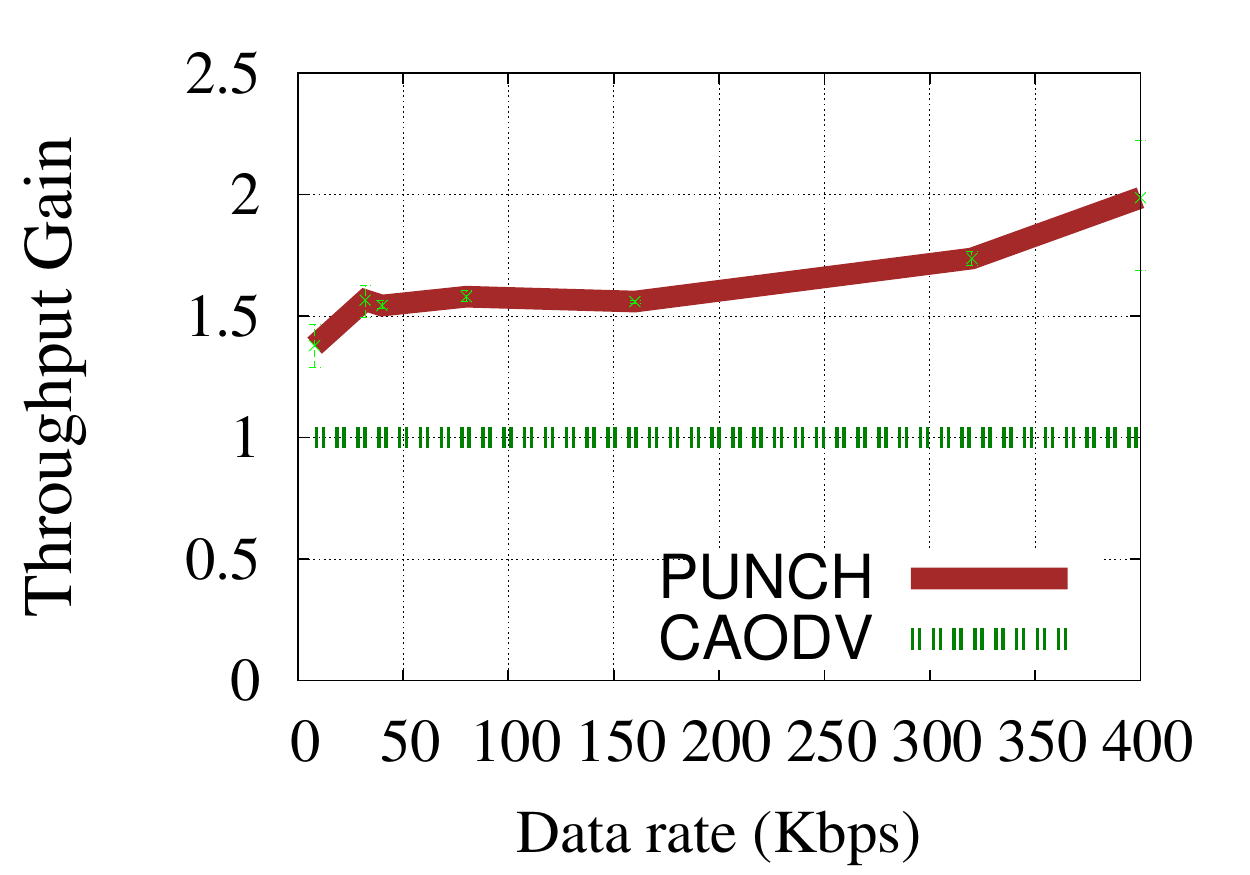}
	\caption{Throughput gain.}
	\label{rate_b}
	\end{subfigure}
	\begin{subfigure}[t]{0.24\textwidth}
\centering
	\includegraphics[width=1.8in]{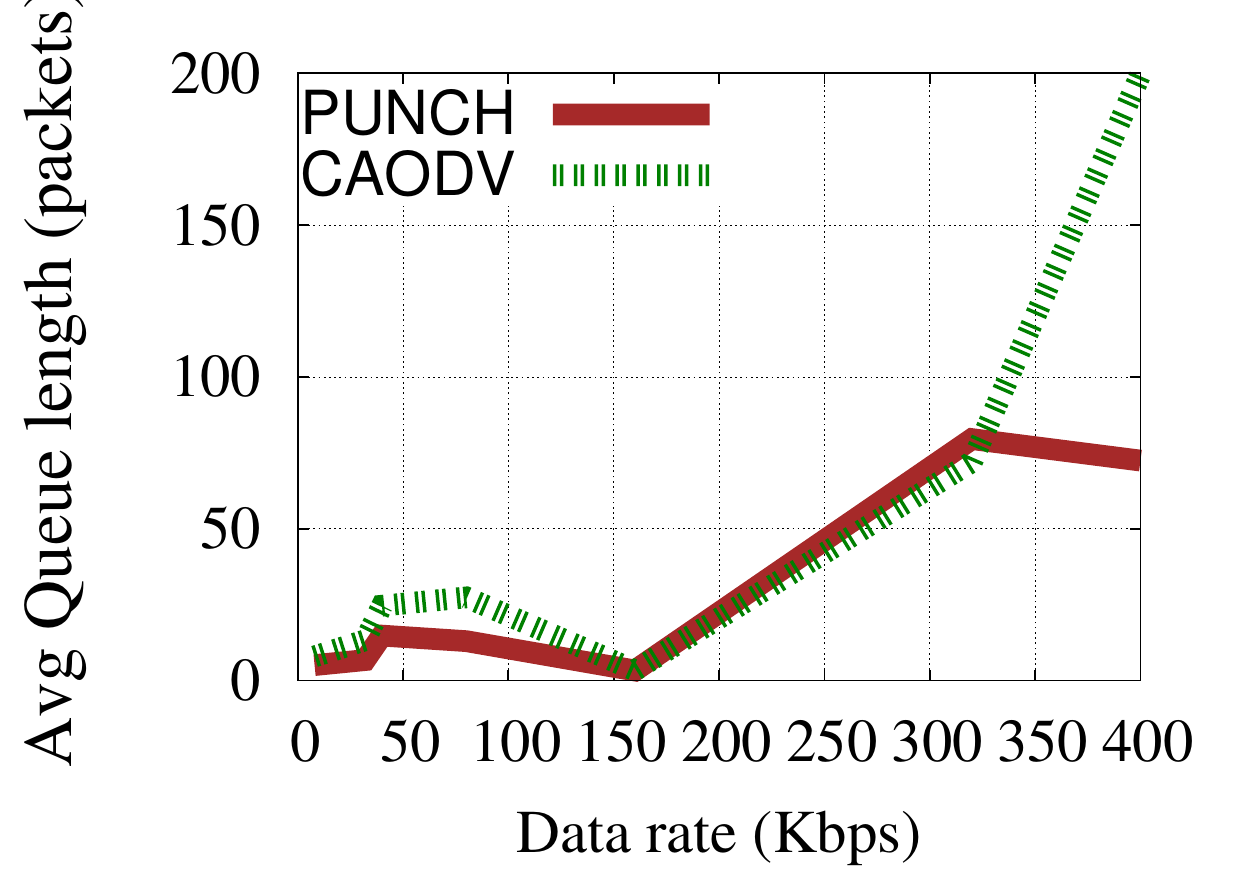}
	\caption{Average queue length.}
	\label{rate_c}
	\end{subfigure}
	\begin{subfigure}[t]{0.24\textwidth}
	\centering
	\includegraphics[width=1.8in]{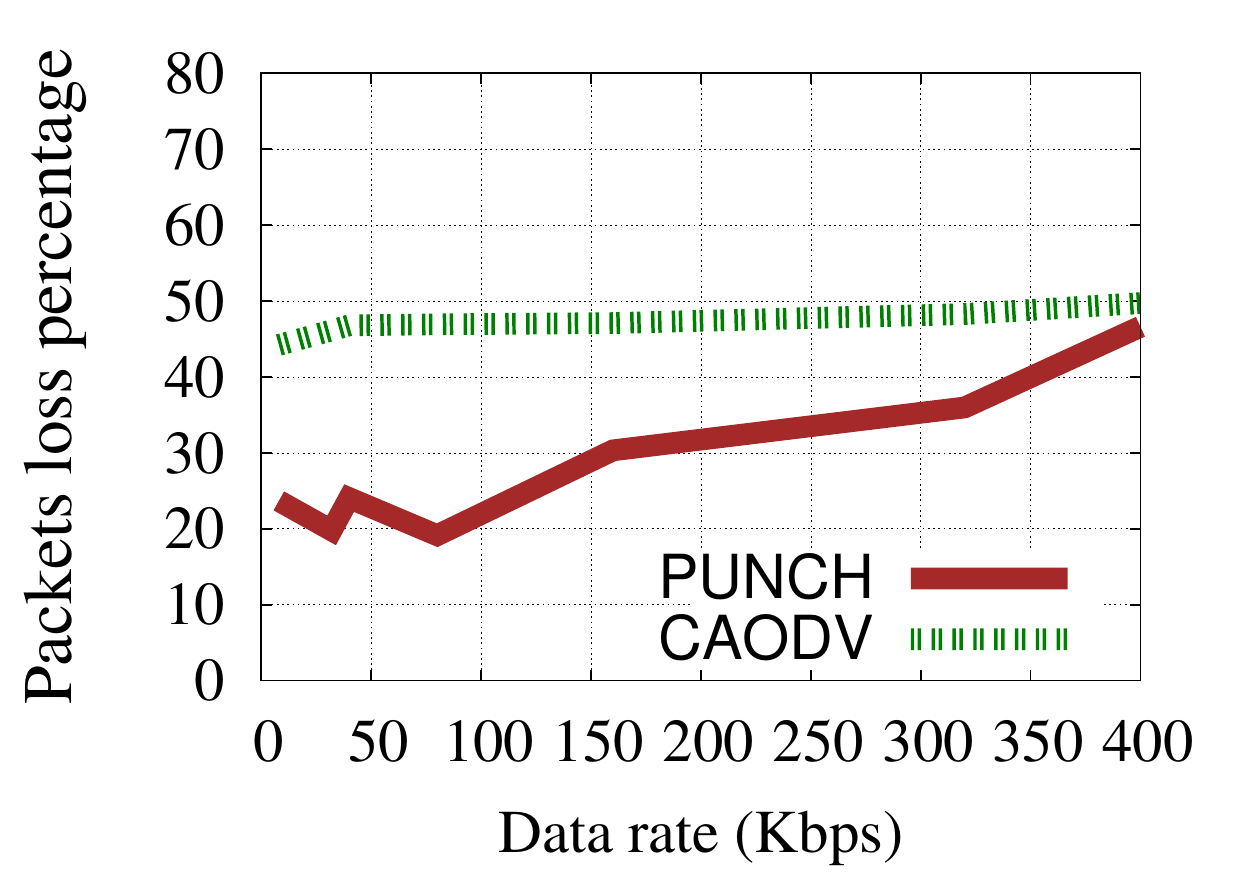}
	\caption{Packet loss percentage.}
	\label{rate_d}
	\end{subfigure}
\caption{Effect of changing the data rate on \sys{} performance compared to CAODV. For the coding and throughout gains, CAODV has a constant performance of one as the gains are normalized relative to it.}
\label{rate}
\end{figure*}

\begin{figure*}[!t]
\centering
	\begin{subfigure}[t]{0.24\textwidth}
	\centering
    \includegraphics[width=1.8in]{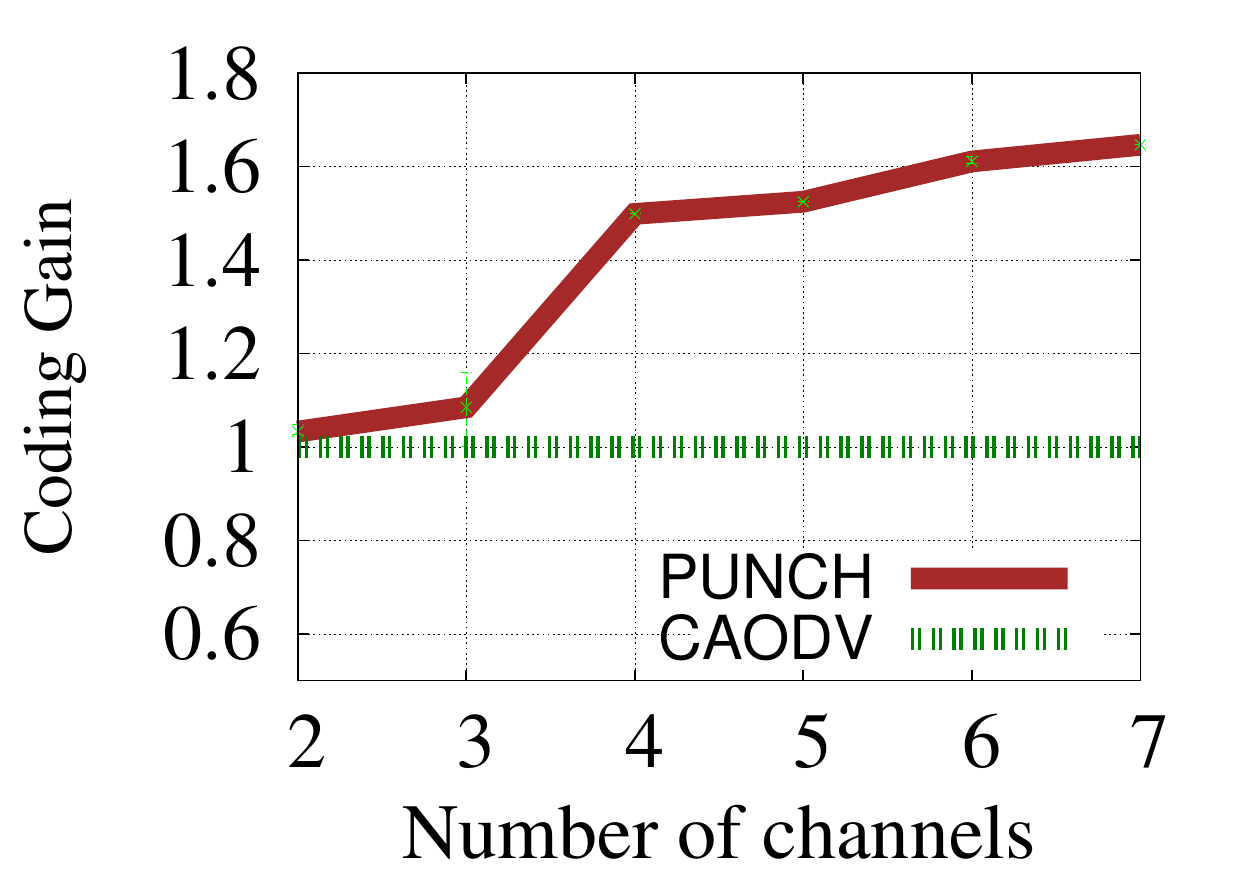}
	\caption{Coding gain.}
	\label{fig:intro_raw}
	\end{subfigure}
	\begin{subfigure}[t]{0.24\textwidth}
	\centering
	\includegraphics[width=1.8in]{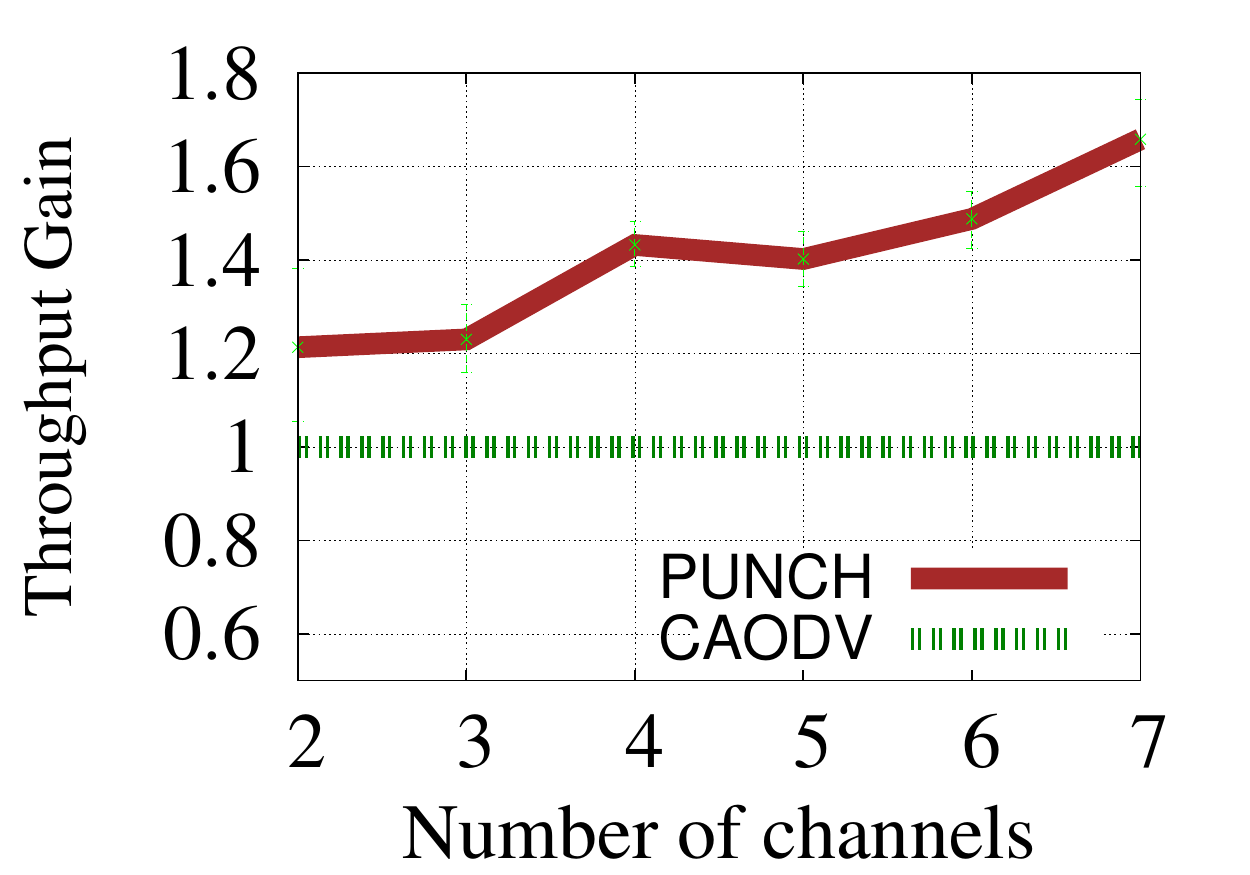}
	\caption{Throughput gain.}
	\label{fig:intro_denoised}
	\end{subfigure}
	\begin{subfigure}[t]{0.24\textwidth}
	\centering
	\includegraphics[width=1.8in]{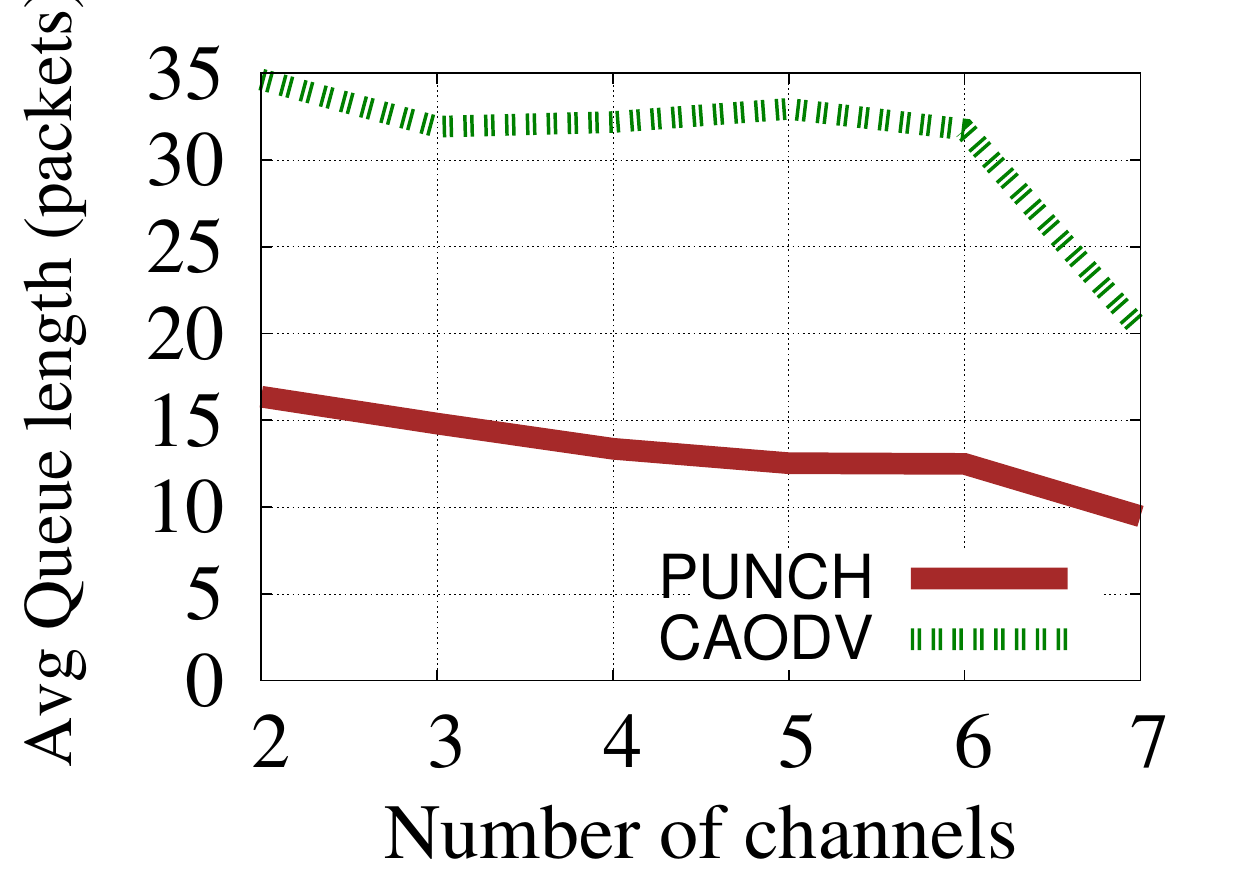}
	\caption{Average queue length.}
	\label{fig:intro_denoised}
	\end{subfigure}
	\begin{subfigure}[t]{0.24\textwidth}
	\centering
	\includegraphics[width=1.8in]{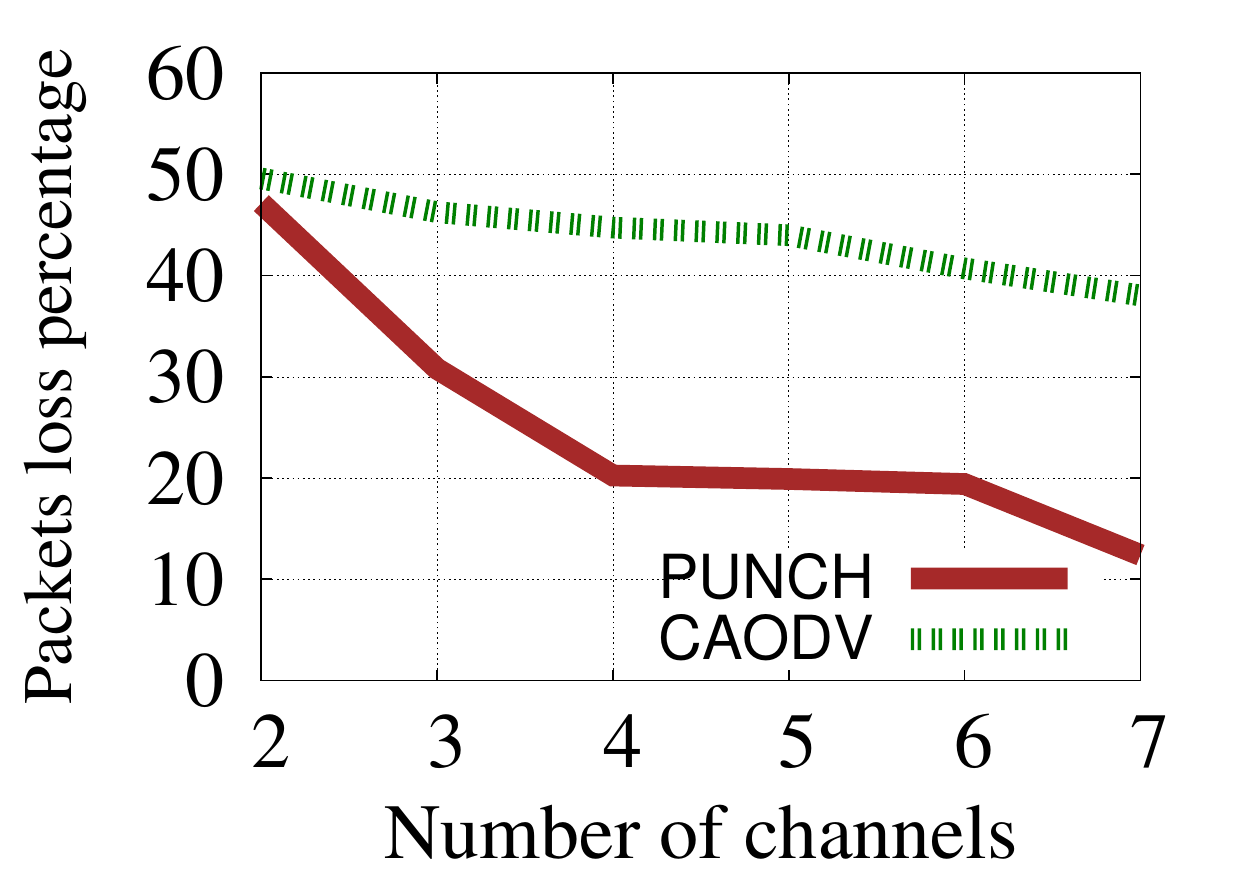}
	\caption{Packet loss percentage.}
	\label{fig:intro_denoised}
	\end{subfigure}
\caption{Effect of changing the number of channels on performance.}
\label{channum}
\end{figure*}

\section{Evaluation}\label{d}
In this section, we evaluate the performance of \sys{} using NS2 simulations. We start by describing the experiment setup. Then, we provide detailed evaluation of the proposed algorithm over different topologies and compare its performance to other CRNs routing protocols.

\subsection{Experimental Testbed}
We used a multi-channel version of NS2 \cite{crextension}.
 We used CAODV \cite{caodv} as the underlying routing protocol we deploy our forwarding algorithm on. CAODV is an extension for the popular AODV protocol in ad hoc networks \cite{Perkins97ad-hocon-demand}  to enable multiple channels and PUs awareness in CRNs. We use the IEEE 802.11 as the MAC protocol. We use the star topology shown in \cref{X} to evaluate the effect of the different parameters on \sys{} performance. A larger random topology is used to compare the performance to CAODV and quantify the advantage of using the PUs' information in the network coding graph construction. We used a CBR traffic model for the generated packets from the SUs. The PUs follow the ON/OFF traffic model described in \cref{a}.

\begin{table}[!t]
\centering
\caption {Default values for the experimental parameters.}
\label{par_table}
\begin{tabular}{|l|c|c|}
\hline
Parameter                          & Default Value & Range    \\ \hline
Data rate (Kbps)                   & 32            & 4 - 200  \\ \hline
Number of channels / node          & 3             & 2 - 7    \\ \hline
Number of PUs                      & 3             & 0 - 4    \\ \hline
PU activity $\lambda$   & 0.5             & 0.25 - 20 \\ \hline
Number of active connections       & 2             & 2        \\ \hline
Packet size (bytes)                & 512           & 512      \\ \hline
\end{tabular}
\end{table}

\subsection{Experiment Parameters}
We evaluate the effect of changing different parameters on the performance of \sys{}. These include the traffic load, number of available channels per node, number of PUs, and activity of the PUs. The default values and the range for each parameter are summarized in \cref{par_table}.

\begin{figure*}[!t]
\centering
	\begin{subfigure}[t]{0.24\textwidth}
	\centering
    \includegraphics[width=1.8in]{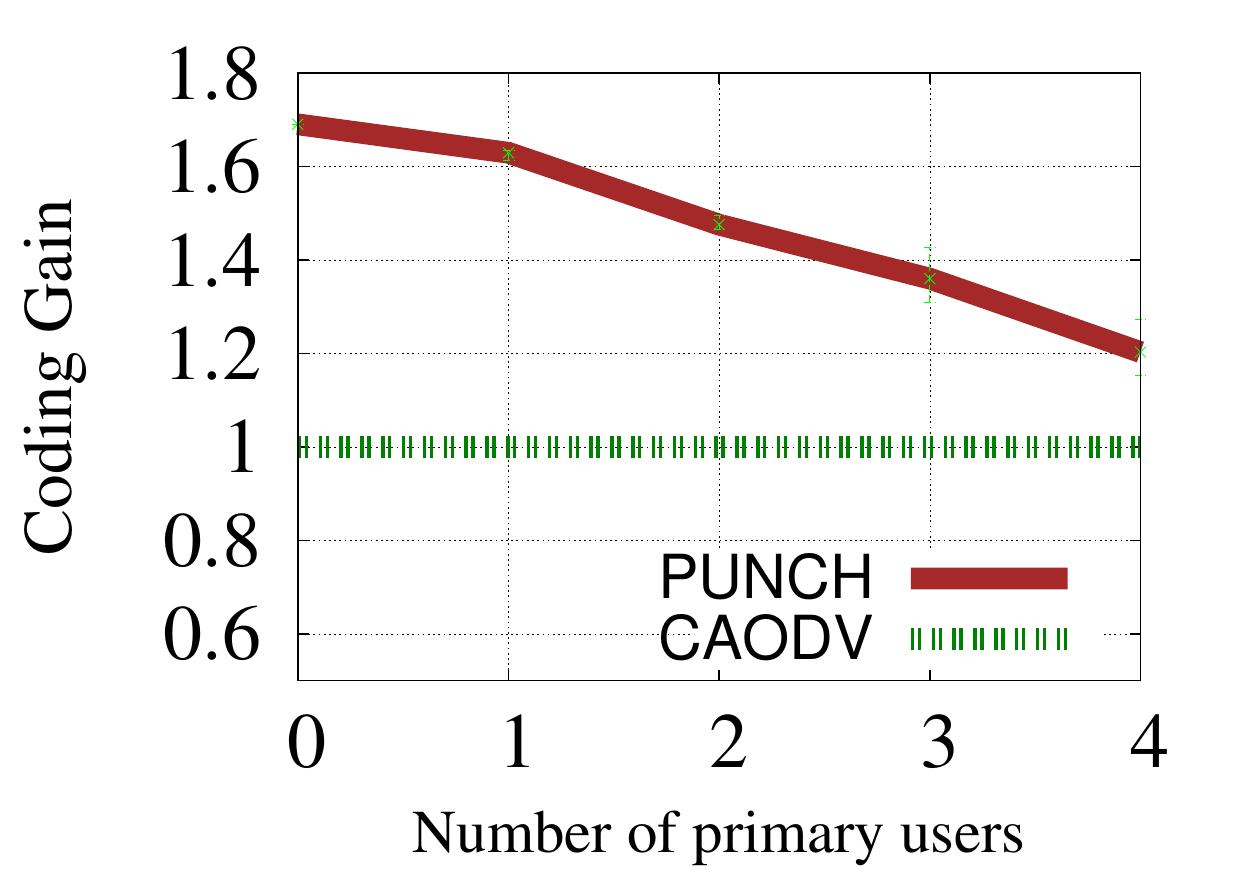}
	\caption{Coding gain.}
	\label{fig:intro_raw}
	\end{subfigure}
	\begin{subfigure}[t]{0.24\textwidth}
	\centering
	\includegraphics[width=1.8in]{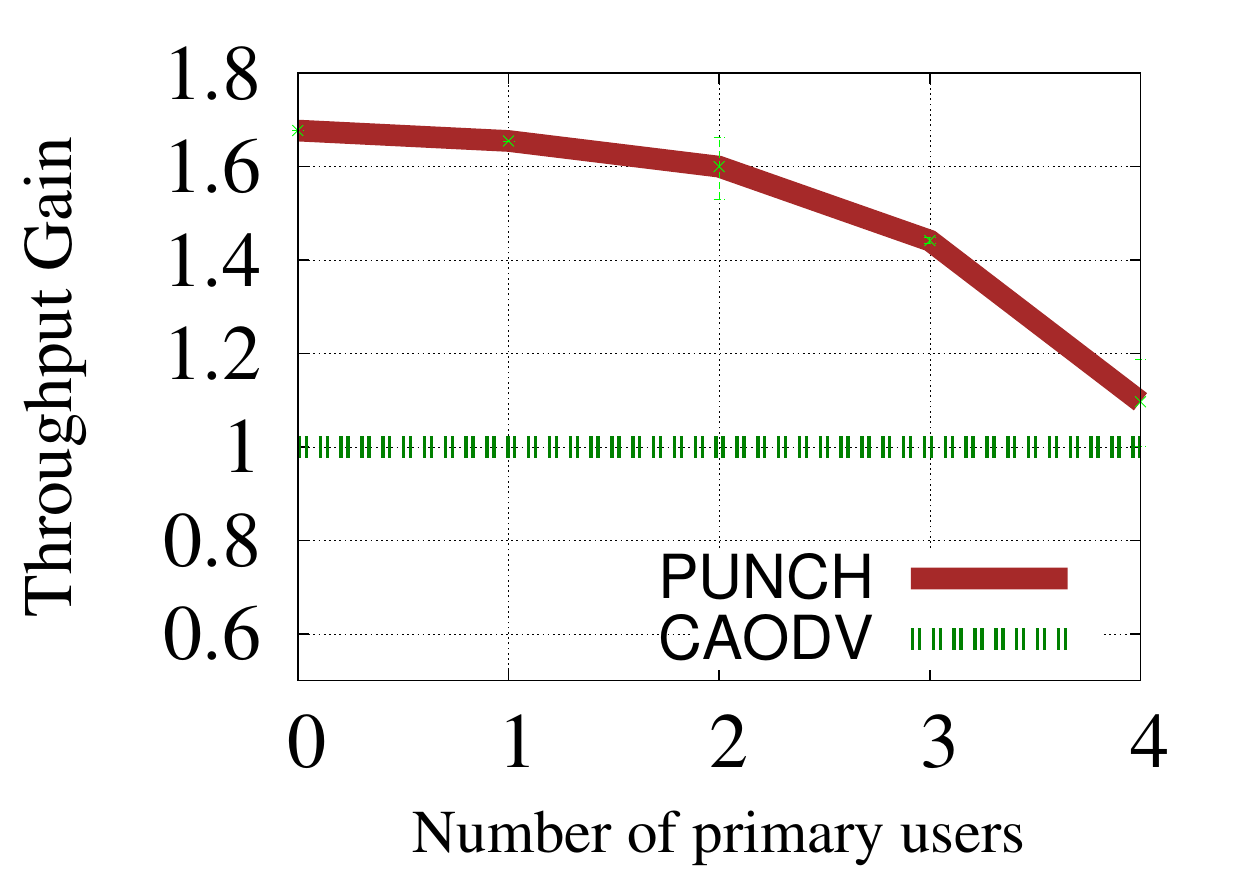}
	\caption{Throughput gain.}
	\label{fig:intro_denoised}
	\end{subfigure}
	\begin{subfigure}[t]{0.24\textwidth}
	\centering
	\includegraphics[width=1.8in]{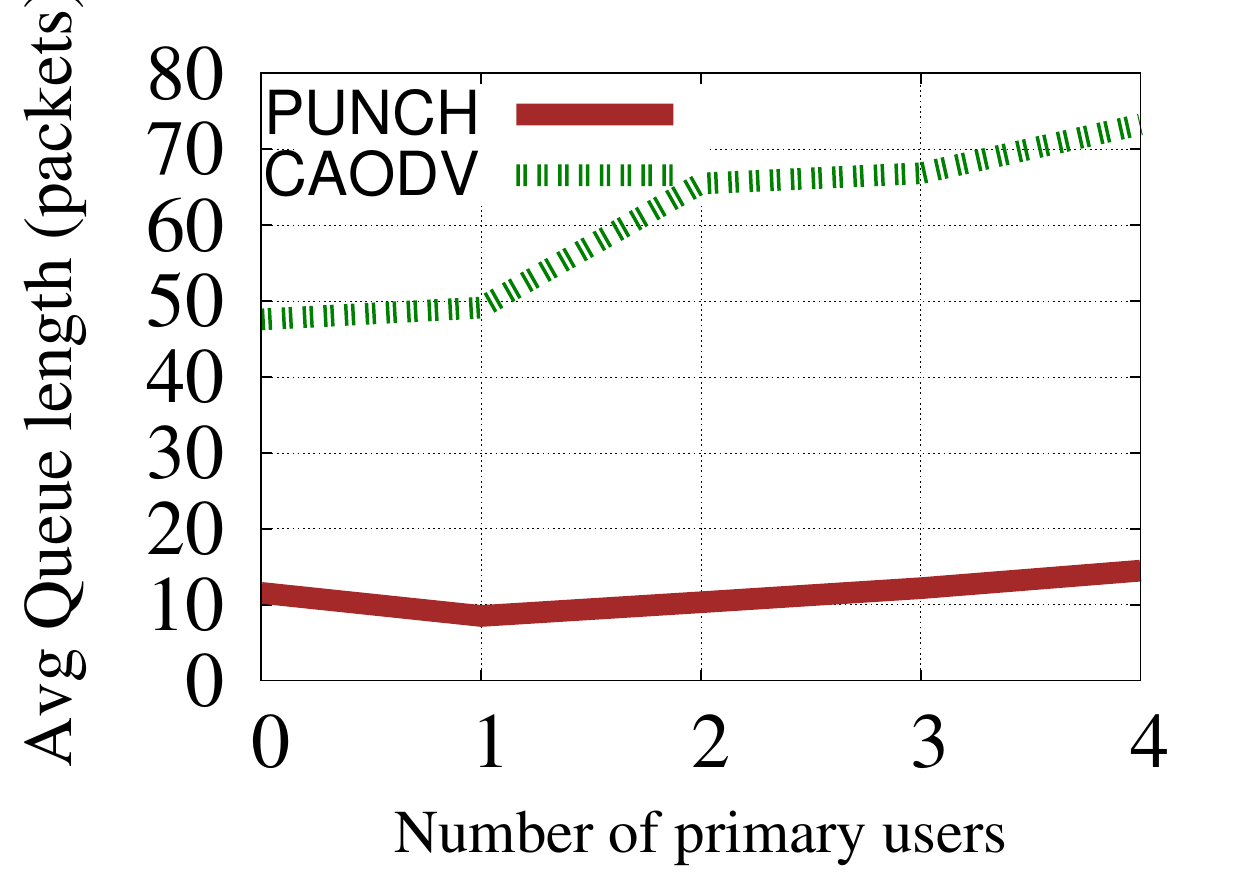}
	\caption{Average queue length.}
	\label{fig:intro_denoised}
	\end{subfigure}
	\begin{subfigure}[t]{0.24\textwidth}
	\centering
	\includegraphics[width=1.8in]{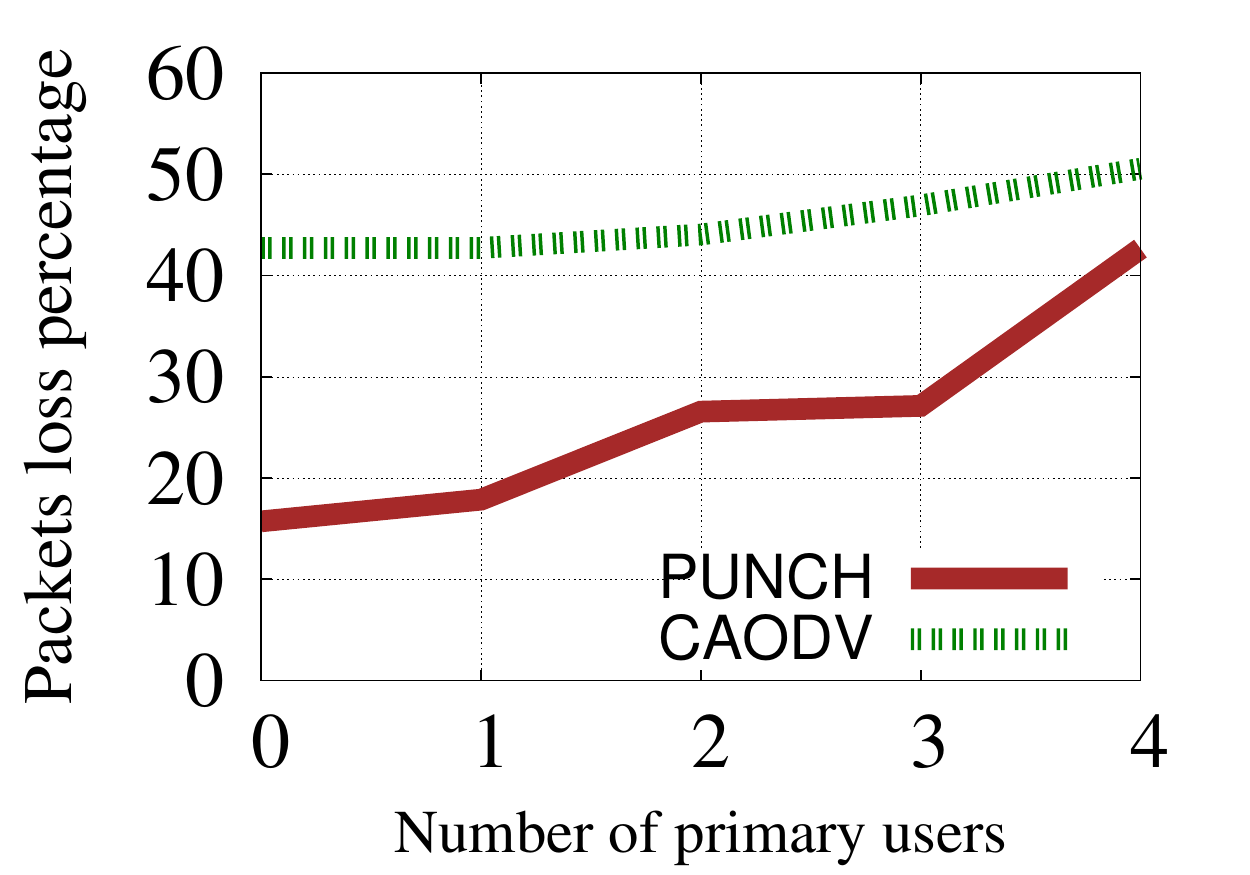}
	\caption{Packet loss percentage.}
	\label{fig:intro_denoised}
	\end{subfigure}
\caption{Effect of changing the number of primary users on performance.}
\label{npu}
\end{figure*}

\begin{figure*}[!t]
\centering
	\begin{subfigure}[t]{0.24\textwidth}
	\centering
    \includegraphics[width=1.8in]{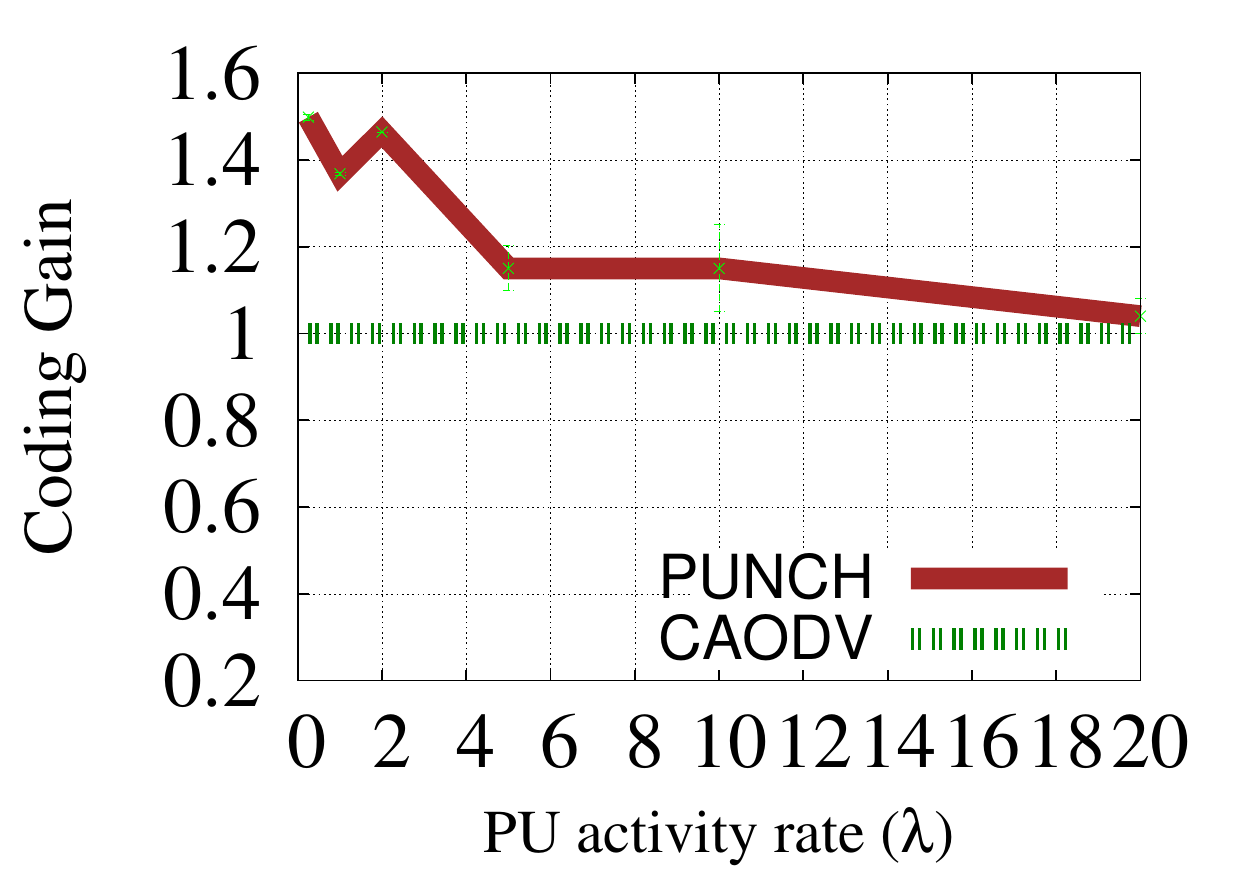}
	\caption{Coding gain.}
	\label{fig:intro_raw}
	\end{subfigure}
	\begin{subfigure}[t]{0.24\textwidth}
\centering
	\includegraphics[width=1.8in]{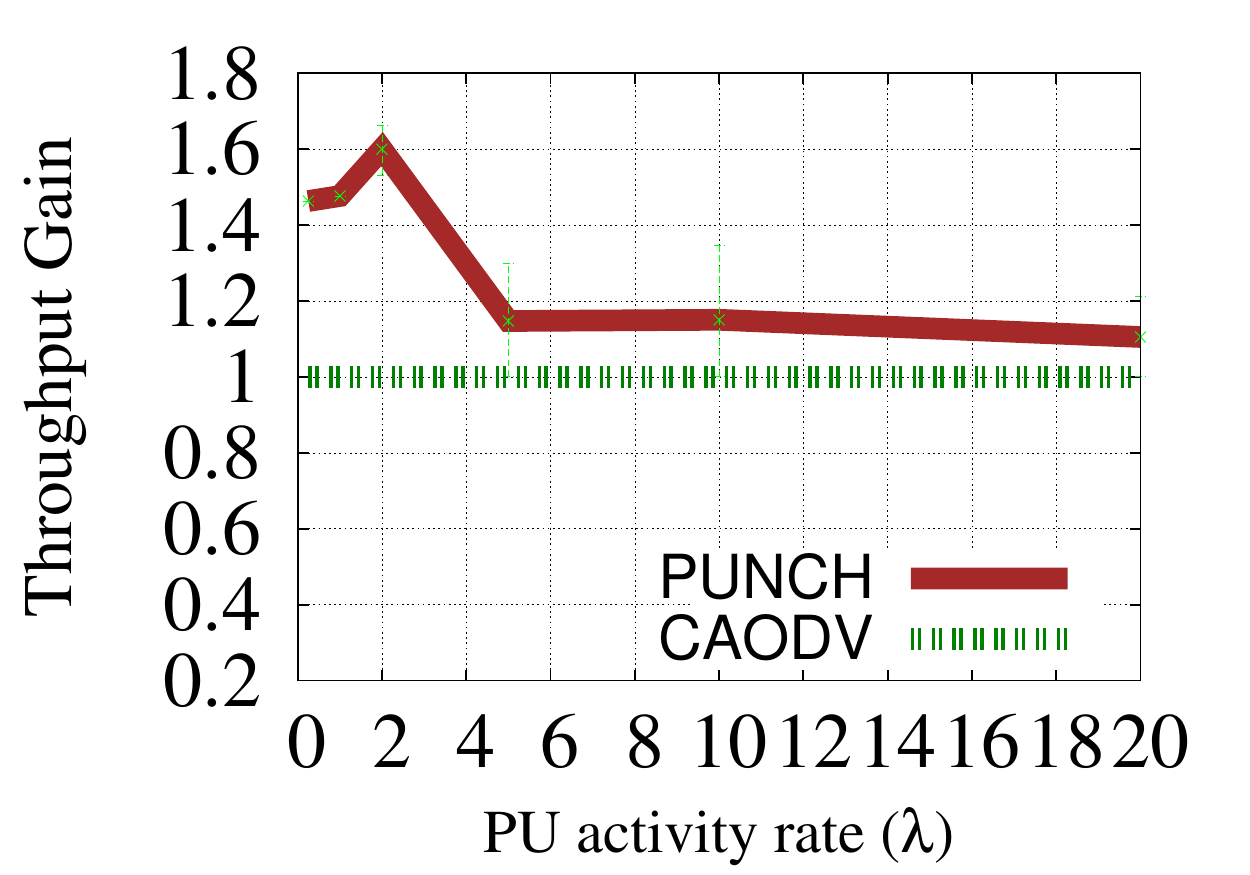}
	\caption{Throughput gain.}
	\label{fig:intro_denoised}
	\end{subfigure}
	\begin{subfigure}[t]{0.24\textwidth}
	\centering
	\includegraphics[width=1.8in]{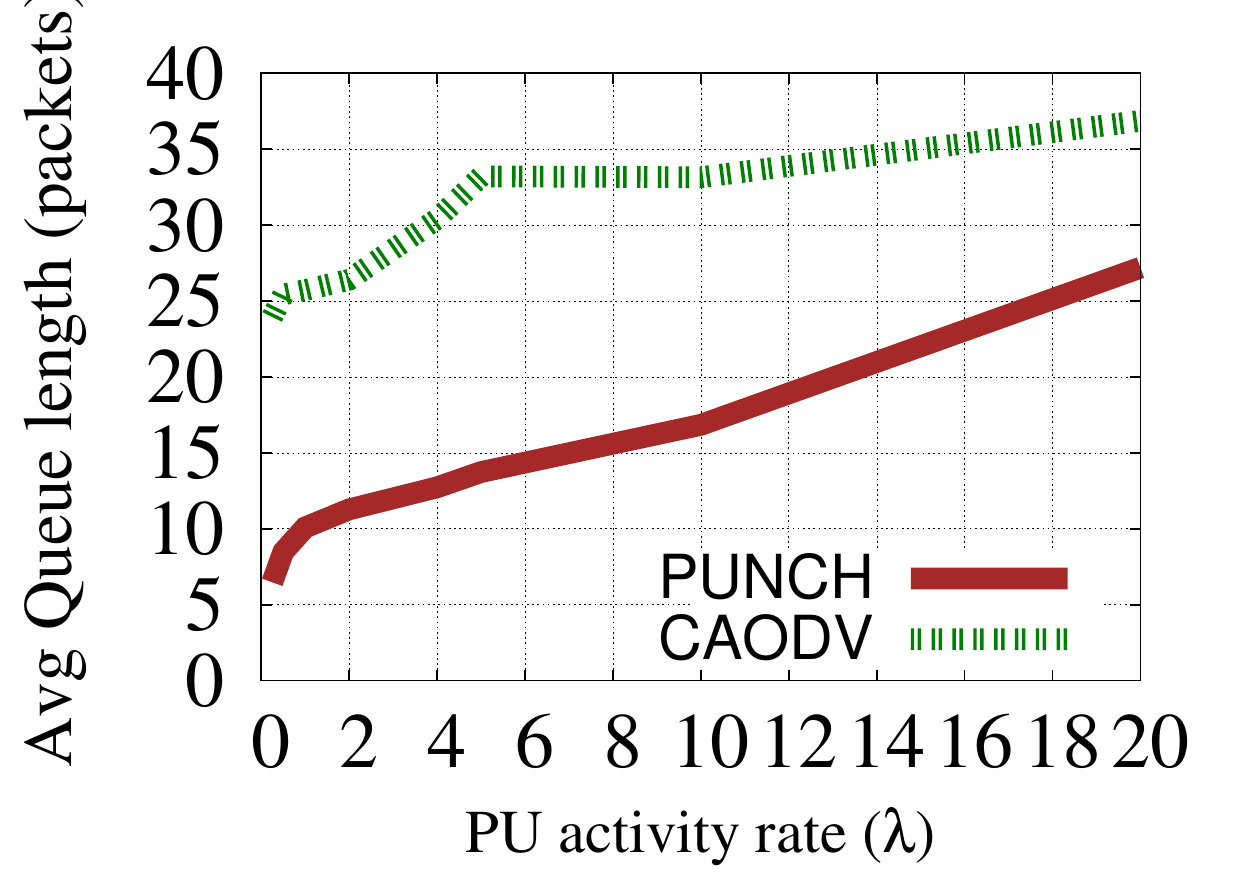}
	\caption{Average queue length.}
	\label{fig:intro_denoised}
	\end{subfigure}
	\begin{subfigure}[t]{0.24\textwidth}
	\centering
	\includegraphics[width=1.8in]{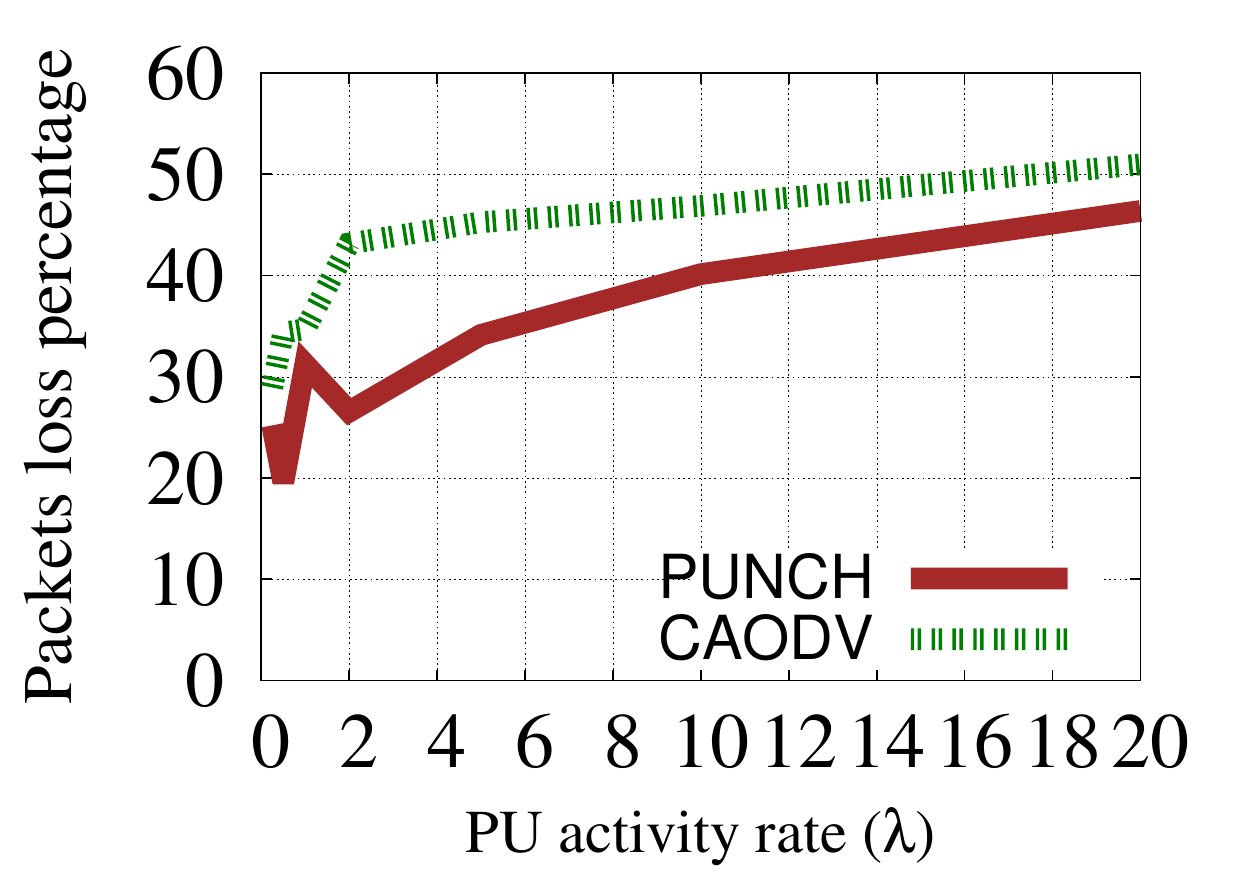}
	\caption{Packet loss percentage.}
	\label{fig:intro_denoised}
	\end{subfigure}
\caption{Effect of changing the primary users' activity rate $\lambda$ on performance.}
\label{activity}
\end{figure*}

\subsection{Metrics}
We also use the following four metrics to evaluate the performance:
\begin{enumerate}
\item Throughput gain: The ratio between the total number of packets that reach their destinations using \sys{} to the total number of packets that reach their destinations without using standard CAODV.
\item Coding gain: The ratio between the total number of transmissions needed without using \sys{} (i.e., using CAODV) to the total number of transmissions needed by \sys{} to deliver the same number of packets ($\ge 1$).
\item Average queue length: The average queue length per node. It captures the effect of network coding on the queue length at each node.
\item Packet loss rate: Is the ratio between the lost packets to the transmitted packets.
\end{enumerate}

\begin{figure*}[!t]
\centering
	\begin{subfigure}[t]{0.24\textwidth}
	\centering
    \includegraphics[width=1.8in]{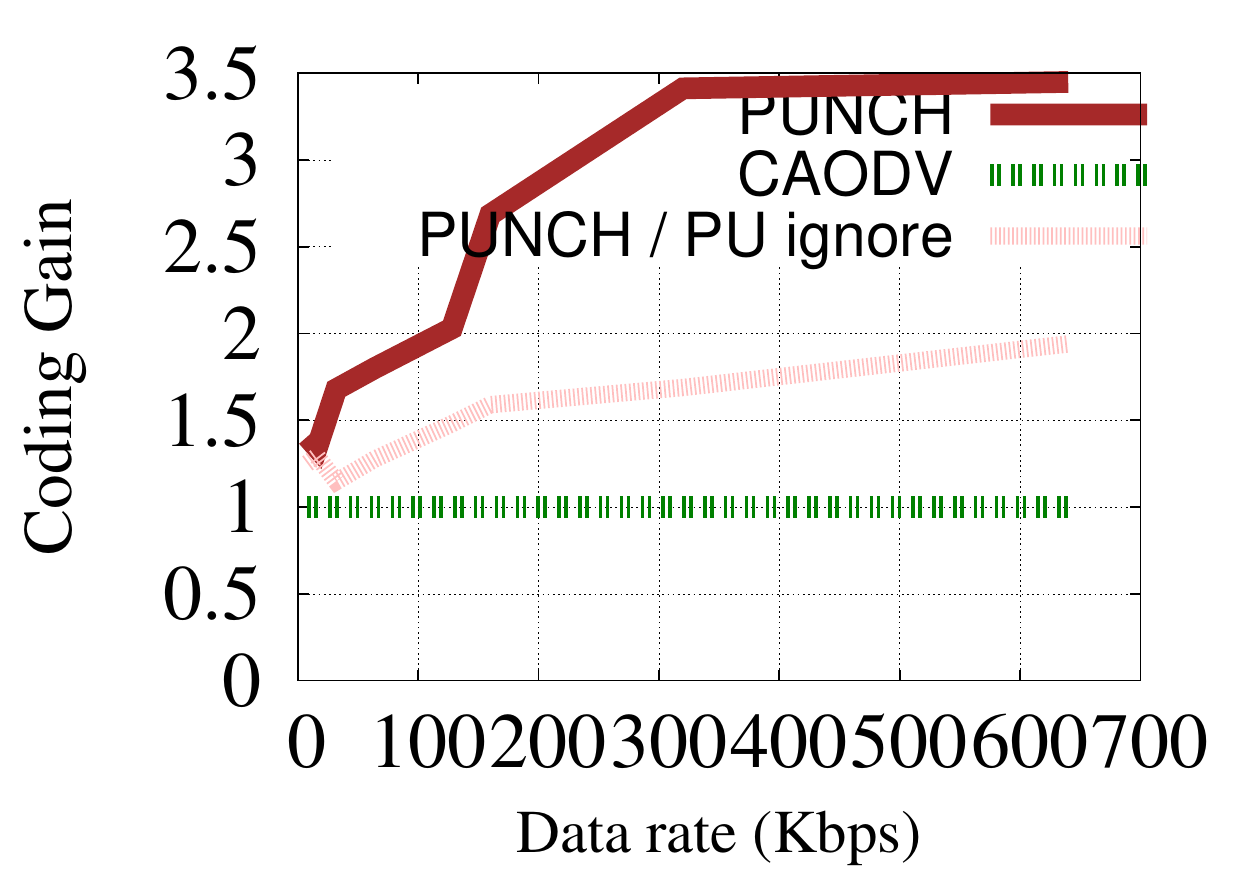}
	\caption{Coding gain.}
	\label{fig:intro_raw}
	\end{subfigure}
	\begin{subfigure}[t]{0.24\textwidth}
	\centering
	\includegraphics[width=1.8in]{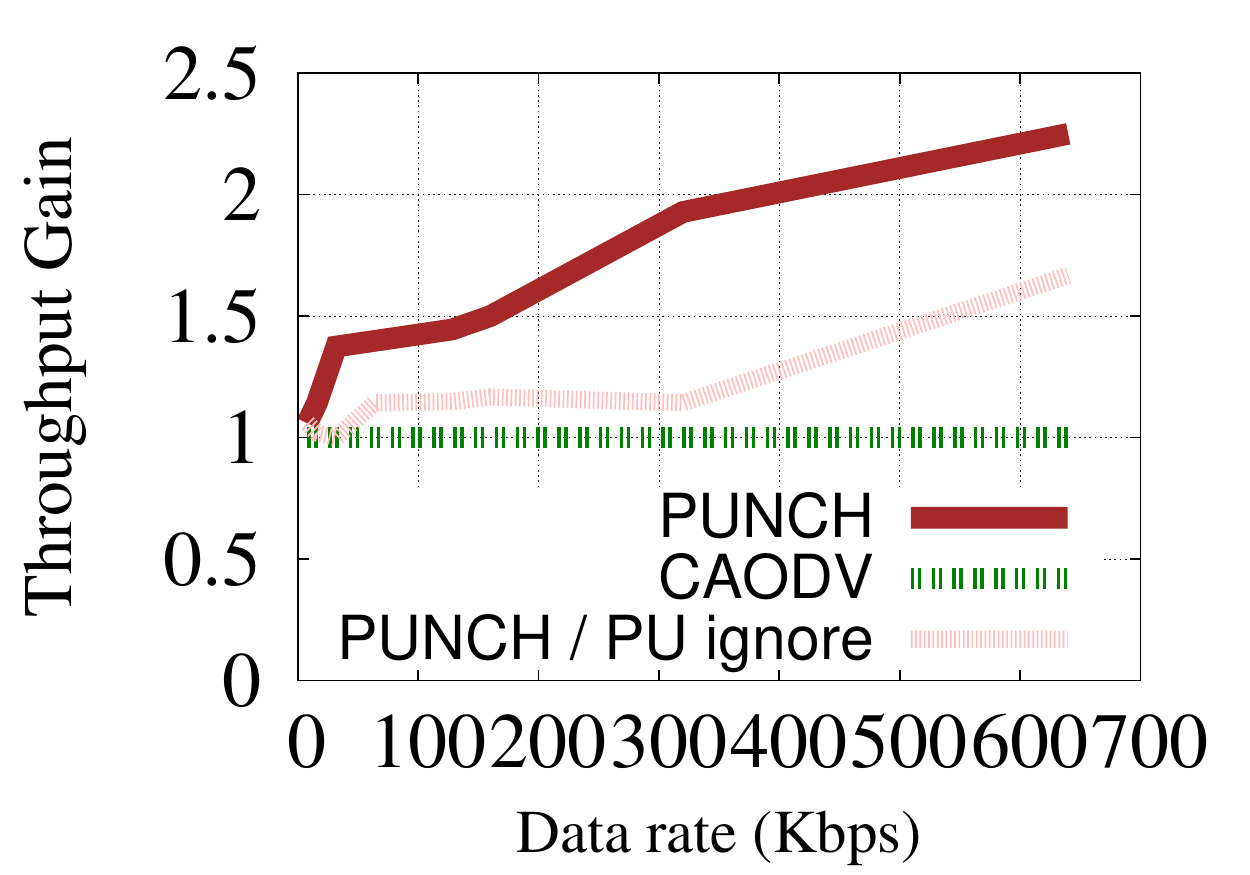}
	\caption{Throughput gain.}
	\label{fig:intro_denoised}
	\end{subfigure}
	\begin{subfigure}[t]{0.24\textwidth}
	\centering
	\includegraphics[width=1.8in]{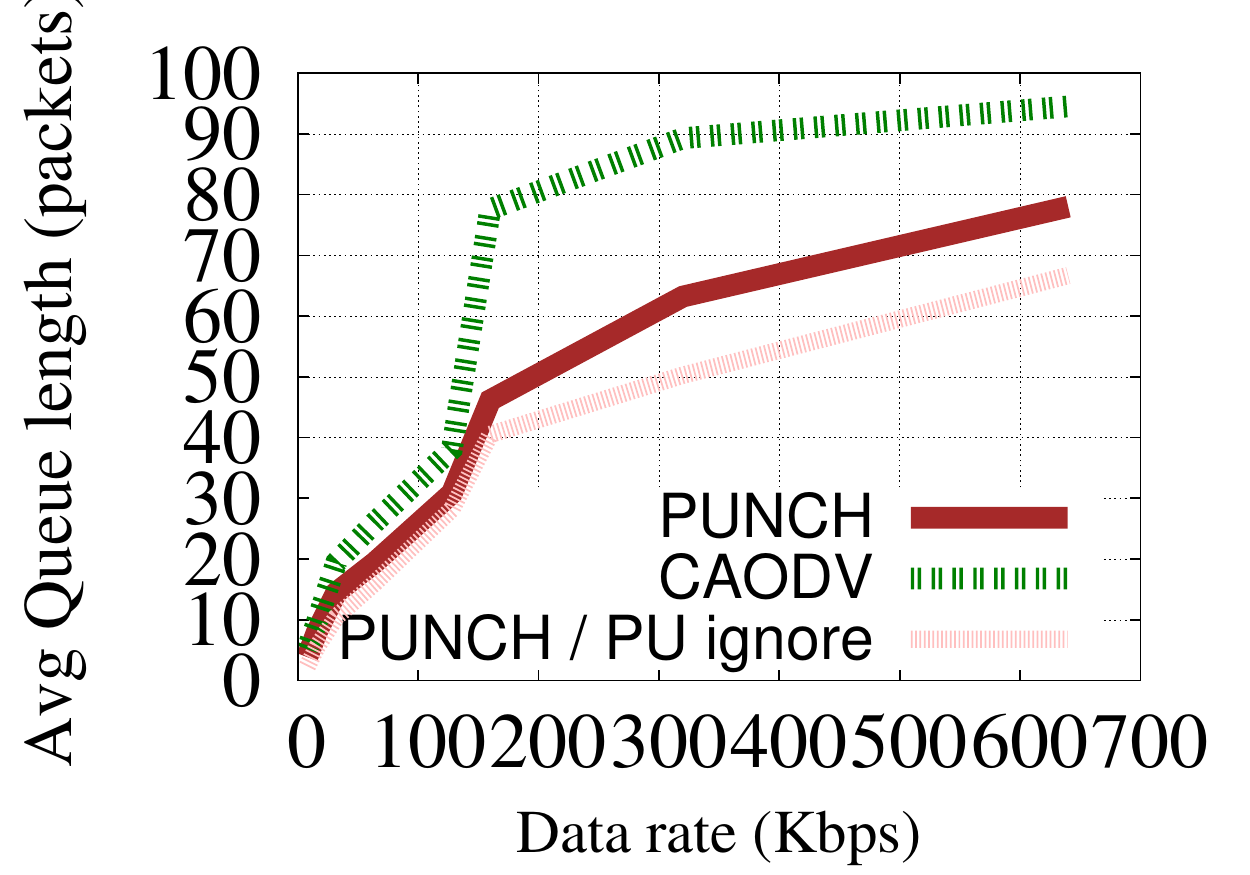}
	\caption{Average queue length.}
	\label{fig:intro_denoised}
	\end{subfigure}
	\begin{subfigure}[t]{0.24\textwidth}
	\centering
	\includegraphics[width=1.8in]{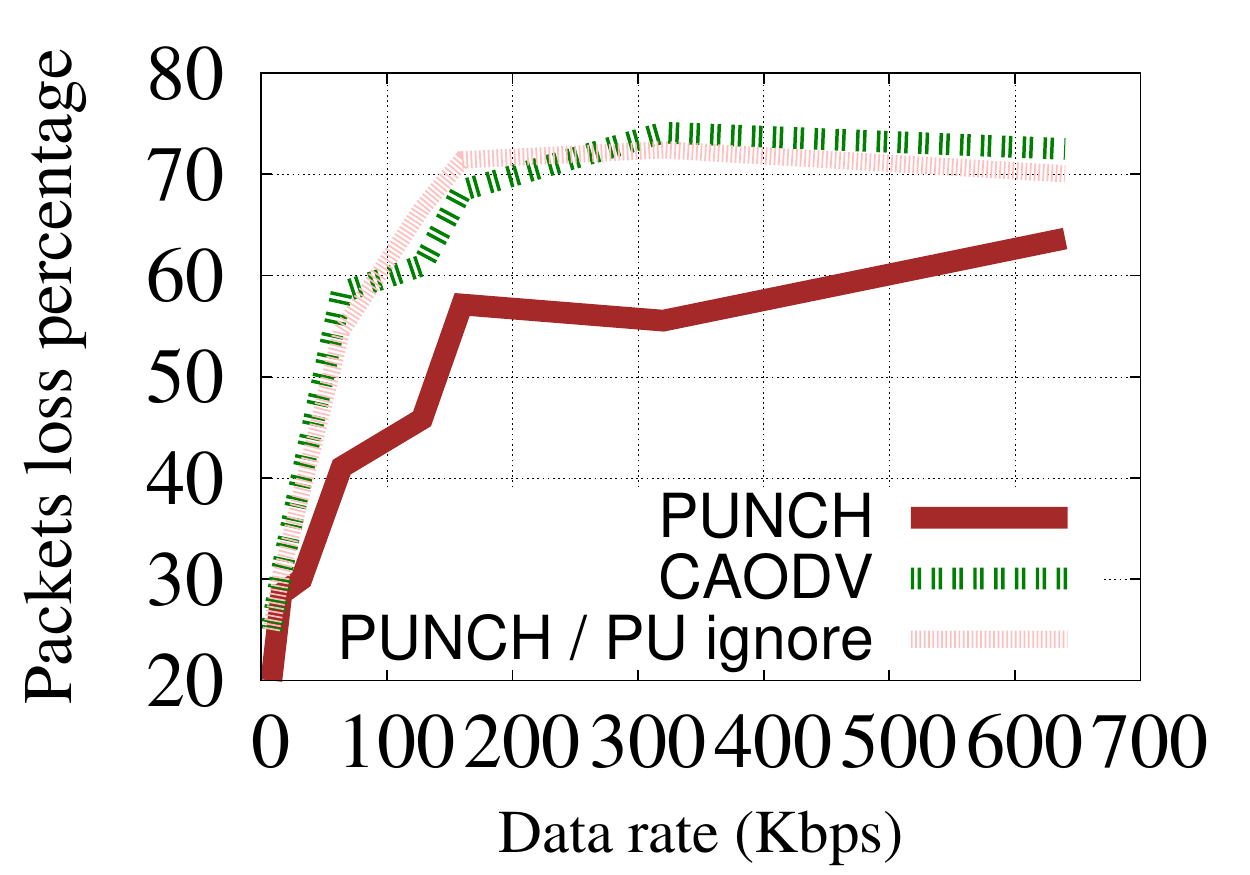}
	\caption{Packet loss percentage.}
	\label{fig:intro_denoised}
	\end{subfigure}
\caption{Performance evaluation for the random topology for the different algorithms.}
\label{random}
\end{figure*}

\subsection{Experimental Results}
\subsubsection{Traffic load}
\cref{rate} shows the effect of increasing the load on the different metrics. The figure shows that increasing load leads to significant increase in both throughput and coding gain for the proposed algorithm as compared to the standard CAODV algorithm\footnote{Note that since we normalize the coding and throughput gains by dividing by that of CAODV, its performance is always constant at 1.}. The figure also shows that, as expected, the average queue length and loss rate both increase with increasing the offered load, with better performance for the proposed algorithm compared to the CAODV algorithm. This can be explained by noting that coding leads to sending more packets per unit time (better throughput and coding gain), and hence empties the queues faster and delivers more packets.

\subsubsection{Number of channels per node}
 The effect of changing the number of channels per node is shown in \cref{channum}. We can notice that increasing the number of channels while fixing the PUs' density leads to better spectrum availability and hence better performance. This is reflected in increasing the coding and throughput gains and reducing the average queue size and loss rate up to the network capacity.

\subsubsection{Number of primary users}
\cref{npu} shows the effect of changing the number of PUs, i.e. increasing the PUs' density, on performance. Increasing the PUs' density reduces the spectrum availability and hence leads to worse performance in terms of reduced throughput and coding gain as well as increased average queue length and loss rate.

\subsubsection{PU activity}
Similarly, increasing the PUs' activity (by increasing the parameter $\lambda$ which reflects the number of packets sent by the PU in a unit time) leads to worse performance.

\subsection{Random Topology}
We also evaluated the performance of the proposed algorithm on a random topology. The topology contains 20 SUs, eight of them are generating packets to random destinations. The rest of the parameters are set to the default values in \cref{par_table}. The results for \sys{}, CAODV, and an algorithm that is similar to \sys{} but without taking PUs' activity into account when constructing the coding graph (PU ignore) are plotted in \cref{random} for different network loads. Other parameters gave similar performance to the star topology results shown in the previous section.

The figure shows that larger networks and more flows lead to even better throughput and coding gains due to the increased opportunities for coding. The algorithm that ignores the PUs' activity can empty the queue faster than \sys{} as it has less constraints on constructing its coding graph. However, the downside is that it has a higher loss ratio than \sys{}. This highlights the importance of using the PUs' activity when choosing the best packet encoding.

\section{Conclusion}\label{e}
We presented \sys{} as a new algorithm of forwarding via network coding in Cognitive Radio Networks (CRNs). \sys{} intelligently mixes packets together to increase throughput and decrease the number of transmissions taking into account both the primary users' activity and the links quality. We formulated the problem as a graph theoretic problem and provided a heuristic for efficiently solving it as well as discussed a number of practical implementation issues.

Evaluation of the proposed algorithm under different scenarios using NS2 simulations showed that \sys{} can increase the throughput of the constrained secondary users' network by 150\% to 200\% for a wide range of scenarios covering different primary users' densities, traffic loads, and spectrum availability. This highlights its capabilities in increasing the spectrum opportunities in CRNs.

Currently, we are expanding \sys{} in different directions including QoS routing in CRNs by modifying the coding graph structure, deploying \sys{} in actual networks, among others.

\ifCLASSOPTIONcompsoc
  \section*{Acknowledgments}
\else

\ifCLASSOPTIONcaptionsoff
  \newpage
\fi

\bibliographystyle{unsrt}

\end{document}